\documentclass[english]{IEEEtran}
\usepackage[T1]{fontenc}
\usepackage[latin9]{inputenc}
\usepackage{float}
\usepackage{amsmath}
\usepackage{amsthm}
\usepackage{amssymb}
\usepackage{graphicx}

\makeatletter

\floatstyle{ruled}
\newfloat{algorithm}{tbp}{loa}
\providecommand{\algorithmname}{Algorithm}
\floatname{algorithm}{\protect\algorithmname}

\theoremstyle{plain}
\newtheorem{thm}{\protect\theoremname}
\theoremstyle{plain}
\newtheorem{lem}[thm]{\protect\lemmaname}

\theoremstyle{definition}
\newtheorem{eg}{Example}

\theoremstyle{plain}
\newtheorem{mythm}{Theorem}

\theoremstyle{definition}
\newtheorem{remark}{Remark}

\makeatother

\usepackage{babel}
\providecommand{\lemmaname}{Lemma}
\providecommand{\theoremname}{Theorem}

\begin{document}

\title{Distributed Stochastic Optimization in Networks with Low Informational
Exchange}

\author{Wenjie~Li and Mohamad~Assaad,~\IEEEmembership{Senior Member, IEEE}
\thanks{This paper was presented in part at 55th Annual Allerton Conference
on Communication, Control, and Computing, Monticello, IL, United States,
Oct 2017 \cite{li2017distributed}.}\thanks{W. Li and M. Assaad are with the Laboratoire des Signaux et Systèmes
(L2S, UMR CNRS 8506), CentraleSupélec, France (e-mail: wenjie.li@lss.centralesupelec.fr;
mohamad.assaad@centralesupelec.fr).}\thanks{W. Li and M. Assaad are also with the TCL Chair on 5G, CentraleSupélec,
France.}}
\maketitle
\begin{abstract}
We consider a distributed stochastic optimization problem in networks
with finite number of nodes. Each node adjusts its action to optimize
the global utility of the network, which is defined as the sum of
local utilities of all nodes. Gradient descent method is a common
technique to solve the optimization problem, while the computation
of the gradient may require much information exchange. In this paper,
we consider that each node can only have a noisy numerical observation
of its local utility, of which the closed-form expression is not available.
This assumption is quite realistic, especially when the system is
too complicated or constantly changing. Nodes may exchange the observation
of their local utilities to estimate the global utility at each timeslot.
We propose stochastic perturbation based distributed algorithms under
the assumptions whether each node has collected local utilities of
all or only part of the other nodes. We use tools from stochastic
approximation to prove that both algorithms converge to the optimum.
The convergence rate of the algorithms is also derived. Although the
proposed algorithms can be applied to general optimization problems,
we perform simulations considering power control in wireless networks
and present numerical results to corroborate our claim.
\end{abstract}


\section{Introduction}

Distributed optimization is a fundamental problem in networks, which
helps to improve the performance of the system by maximizing some
predefined objective function. Significant amount of work have been
done to solve the optimization problems in various applications. For
example, in power control \cite{chiang2008power,leung2006opportunistic,hassan2013distributed}
and beamforming allocation \cite{rashid1998joint,matskani2008convex,wang2015joint,sheng2017joint}
problems, transmitters need to control its transmission power or beamforming
in a smart manner, in order to maximize some performance metric of
the wireless communication systems, such as throughput or energy efficiency.
In medium access control problem \cite{chen2010random}, users set
their individual channel access probability to maximize their benefit.
In wireless sensor networks, sensor nodes collect information to serve
a fusion center, an interesting problem is to make each node independently
decide the quality of its report to maximize the average quality of
information gathered by the fusion center subject to some power constraint
\cite{liu2012optimizing,neely2016distributed}, note that a higher
level of quality requires higher power consumption.

This paper considers an optimization problem in a distributed network
where each node adjusts its own action to maximize the global utility
of the system, which is also perturbed by a stochastic process, \emph{e.g}.,
wireless channels. The global utility is the sum of the local utilities
of all nodes of the network. Gradient descent method is the most common
technique to deal with optimization problems. In many scenarios in
practice, however, the computation of gradient may require too much
information exchange between the nodes, examples are provided in Section~\ref{sec:Motivating-Example}.
Furthermore, there are other contexts also where the utility function
of each node does not have a closed form expression or the expression
is very complex which makes it very hard to use in the optimization,
\emph{e.g.}, computation of the derivatives is very complicated or
not possible. In this paper, we consider therefore that a node only
has a noisy numerical observation of its utility function, which is
quite realistic when the system is complex and time-varying. The nodes
can only exchange the observation of their local utilities so that
each node can have the knowledge of the whole network. However, a
node may not receive all the local utilities of the other nodes due
to the network topology or other practical issues, \emph{e.g.}, it
is not possible to exchange much signaling information. In this situation,
a node has to approximate the global utility with only incomplete
information of local utilities. We have also taken in account such
issue in this paper.

In summary, our problem is quite challenging due to the following
reasons: \emph{i}) each node has only a numerical observation of its
local utility at each time; \emph{ii}) each node may have incomplete
information of the global utility of the network; \emph{iii}) the
action of each node has an impact on the utilities of the other nodes
in the network; \emph{iv}) the utility of each node is also influenced
by some stochastic process (\emph{e.g.}, time varying channels in
wireless networks) and the objective function is the average global
utility. 

In this paper, we develop novel distributed algorithms to optimize
the global average utility of a network, where the nodes can only
exchange the numerical observation of their local utility. Different
versions of algorithms are proposed depending on: \emph{i}) whether
the value of action is constrained or unconstrained; \emph{ii}) whether
each node has the full knowledge of local utilities of all the other
nodes or only a part of the utilities of other nodes. We have proved
the convergence of the algorithms in all situations, using stochastic
approximation tools. The convergence rate of different algorithms
are also derived, in order to show the convergence speed of the proposed
algorithms to the optimum from a quantitative point of view and deeply
investigate the impact of the parameters introduced by the algorithm.
Our theoretical results are further justified by simulations.

Some preliminary results of our work have been presented in \cite{li2017distributed}.
This extended version provides the complete proof of all the results.
Moreover, the constrained optimization problem and the analysis of
the convergence rate in this paper are not considered at all in \cite{li2017distributed}.
As we will see in Section~\ref{sec:Convergence-rate}, the derivation
of the convergence rate is especially challenging.

The rest of the paper is organized as follows. Section~\ref{sec:Related-Work}
discusses some related work and highlights our main contribution.
Section~\ref{sec:System-Model} describes the system model as well
as some basic assumptions. Section~\ref{sec:Motivating-Example}
shows motivating examples to explain the interest of our problem.
Section~\ref{sec:Distributed-allocation-algorithm} develops the
initial version of our distributed optimization algorithm using stochastic
perturbation (DOSP) and shows its convergence. Section~\ref{sec:ESSP-Algorithm-incomplete}
presents the first variant of the DOSP algorithm to deal with the
situation where a node has incomplete information of the global utility
of the network. Section~\ref{sec:constrained} proposes the second
variant of the DOSP algorithm to solve the constrained optimization
problem. Section~\ref{sec:Convergence-rate} focuses on the analysis
of the convergence rate of the proposed algorithms. Section~\ref{sec:Simulation-Results}
shows some numerical results as well as a comparison with an alternative
algorithm and Section~\ref{sec:Conclusion} concludes this paper.

\section{Related Work\label{sec:Related-Work}}

Most of the prior work in the area of optimization consider that the
objective function has a well known and simple closed form expression.
Under this assumption, the optimization problem can be performed using
gradient ascent or descent method \cite{snyman2005practical}. This
method can achieve a local optimum or global optimum in some special
cases (\emph{e.g}. concavity of the utility, etc.) of the optimization
problem. A distributed asynchronous stochastic gradient optimization
algorithms is presented in \cite{Bertsekas1986}. Incremental sub-gradient
methods for non-differentialable optimization are discussed in \cite{Nedic2001}.
Interested readers are referred to a survey by Bertsekas \cite{Bertsekas2010}
on incremental gradient, sub-gradient, and proximal methods for convex
optimization. The use of gradient-based method supposes in advance
that the gradient can be computed or is available at each node, which
is not always possible as this would require too much information
exchanges. In our case, the computation of the gradient is not possible
at each node since only limited control information can be exchanged
in the network. This problem is known as derivative-free optimization,
see \cite{rios2013derivative} and the references therein. Our goal
is then to develop an algorithm that requires only the knowledge of
a numerical observation of the utility function. The obtained algorithm
should be distributed. 

Distributed optimization has also been studied in the literature using
game theoretic tools. However, most of the existing work assume that
a closed form expression of the payoff is available. One can refer
to \cite{bennis2013self,lasaulce2011game} and the references therein
for more details, while we do not consider non-cooperative games in
this paper.

Stochastic approximation (SA) \cite{borkar2008stochastic,kushner2012stochastic}
is an efficient method to solve the optimization problems in noisy
environment. Typically, the action is updated as follows 
\begin{equation}
\boldsymbol{a}_{k+1}=\boldsymbol{a}_{k}+\beta_{k}\widehat{\boldsymbol{g}}_{k}.\label{eq:algo_update}
\end{equation}
where $\widehat{\boldsymbol{g}}_{k}$ represents an estimation of
the gradient $\boldsymbol{g}_{k}$. An important assumption is that
the estimation error $\boldsymbol{\boldsymbol{\varepsilon}}_{k}=\widehat{\boldsymbol{g}}_{k}-\boldsymbol{g}_{k}$
is seen as a zero-mean random vector with finite variance, for example,
see \cite{mertikopoulos2016distributed}. If the step-size $\beta_{k}$
is properly chosen, then $\boldsymbol{a}_{k}$ can tend to its optimum
point asymptotically. The challenge of our work is how to propose
such estimation of the gradient only with the noisy numerical observation
of the utilities.

Most of the previous work related to derivative-free optimization
consider a control center that updates the entire action vector during
the algorithm, see \cite{rios2013derivative} for more details. However,
in our distributed setting, each node is only able to update its own
action. Nevertheless, a stochastic approximation method using the
simultaneous perturbation gradient approximation (SPGA) \cite{spall1992multivariate}
can be an option to solve our distributed derivative-free optimization
problem. The SPGA algorithm was initially proposed to accelerate the
convergence speed of the centralized multi-variate optimization problem
with deterministic objective function. Two measurements of the objective
function are needed per update of the action. The approximation of
the partial derivative of an element~$i$ is given by 
\begin{equation}
\widehat{g}_{i,k}=\frac{f\left(\boldsymbol{a}_{k}+\gamma_{k}\boldsymbol{\Delta}_{k}\right)-f\left(\boldsymbol{a}_{k}-\gamma_{k}\boldsymbol{\Delta}_{k}\right)}{2\gamma_{k}\Delta_{i,k}},\label{eq:g_est2}
\end{equation}
where $\gamma_{k}>0$ is vanishing and $\boldsymbol{\Delta}_{k}=\left[\Delta_{1,k},\ldots,\Delta_{N,k}\right]$
with each element $\Delta_{i,k}$ zero mean and i.i.d. Two successive
measurements of the objective function are required to perform a single
estimation of the gradient. The interest of the SPGA method is that
each variable can be updated simultaneously and independently. Spall
has also proposed an one-measurement version of the SPGA algorithm
in \cite{spall1997one} with 
\begin{equation}
\widehat{g}_{i,k}=\frac{f\left(\boldsymbol{a}_{k}+\gamma_{k}\boldsymbol{\Delta}_{k}\right)}{2\gamma_{k}\Delta_{i,k}}.\label{eq:g_est2-1}
\end{equation}
Such algorithm also leads $\boldsymbol{a}_{k}$ to converge, while
with a decreased speed compared with the two-measurement SPGA. An
essential result is that the estimation of gradient using (\ref{eq:g_est2})
or (\ref{eq:g_est2-1}) is unbiased if $\gamma_{k}$ is vanishing,
as long as the objective function $f$ is static. However, if the
objective function is stochastic and its observation is noisy, there
would be an additional term of stochastic noise in the numerator of
(\ref{eq:g_est2}) and (\ref{eq:g_est2-1}), which may seriously affect
the performance of approximation when the value $\gamma_{k}$ is too
small. As a consequence, the SPGA algorithm cannot be used to solve
our stochastic optimization problem.

The authors in \cite{frihauf2012nash} proposed a fully distributed
Nash equilibrium seeking algorithm which requires only a measurement
of the numerical value of the \emph{static} utility function. Their
scheme is based on deterministic sine perturbation of the payoff function
in continuous time. In \cite{hanif2012convergence}, the authors extended
the work in \cite{frihauf2012nash} to the case of discrete time and
stochastic state-dependent utility functions, convergence to a close
region of Nash equilibrium has been proved. However, in a distributed
setting, it is challenging to ensure that the sine perturbation of
different nodes satisfy the orthogonality requirement, especially
when the number of nodes is large. Moreover, the continuous sine perturbation
based algorithm converges slowly in a discrete-time system. Stochastic
perturbation based algorithm has been proposed in \cite{manzie2009extremum}
to solve an optimization problem, the algorithm is given by 
\begin{equation}
\boldsymbol{a}_{k+1}=\boldsymbol{a}_{k}+\beta\boldsymbol{v}_{k}f\left(\boldsymbol{a}_{k}+\boldsymbol{v}_{k}\right),\label{eq:algo_update-1}
\end{equation}
with $\boldsymbol{v}_{k}$ the zero-mean stochastic perturbation.
The behavior of (\ref{eq:algo_update-1}) has been analyzed in \cite{manzie2009extremum},
however, under the assumption that the objective function is static
and quadratic. Our proposed algorithm is different from (\ref{eq:algo_update-1})
as we use the random perturbation with vanishing amplitude. In addition,
the objective function is stochastic with non-specified form in our
setting, which is much more challenging. Furthermore, we consider
a situation where nodes have to exchange their local utilities to
estimate the global utility and each node may have incomplete information
of the local utilities of other nodes.

\section{System Model\label{sec:System-Model}}

This section presents the problem formulation as well as the basic
assumptions. Throughout this paper, matrices and vectors are in boldface
upper-case letters and in in bold-face lower-case letters respectively.
Calligraphic font denotes set. $\left\Vert \boldsymbol{a}\right\Vert $
denotes the Euclidean norm of any vector $\boldsymbol{a}$. In order
to lighten the notations, we use
\[
F_{i}'\left(\boldsymbol{a}\right)=\frac{\partial F}{\partial a_{i}}\left(\boldsymbol{a}\right),\qquad F_{i,j}''\left(\boldsymbol{a}\right)=\frac{\partial^{2}F}{\partial a_{i}\partial a_{j}}\left(\boldsymbol{a}\right),
\]
and $\nabla F\left(\boldsymbol{a}\right)=\left[F_{i}'\left(\boldsymbol{a}\right),\ldots,F_{N}'\left(\boldsymbol{a}\right)\right]$
in the rest of the paper.

\subsection{Problem formulation}

Consider a network consisting of a finite set of nodes $\mathcal{N}=\left\{ 1,\ldots,N\right\} $.
Each node~$i$ is able to control its own action $a_{i,k}$ at each
discrete timeslot~$k$, in order to maximize the performance of the
network. Introduce the action vector $\boldsymbol{a}_{k}=\left[a_{1,k},\ldots,a_{N,k}\right]^{T}$
which contains the action of all nodes at timeslot~$k$. Let $\mathcal{A}_{i}$
denote the feasible action set of node~$i$, \emph{i.e.}, $a_{i,k}\in\mathcal{A}_{i}$.
Introduce $\mathcal{A}=\mathcal{A}_{1}\times\ldots\times\mathcal{A}_{N}$.
In general, the performance of a network is not only determined by
the action of nodes, but also affected by the environment state, \emph{e.g.},
channels. We assume that the environment state of the entire network
at any timeslot~$k$ is described by a matrix $\mathbf{S}_{k}\in\mathcal{S}$,
which is considered as an i.i.d. ergodic stochastic process. 

For any realization of $\boldsymbol{a}$ and $\mathbf{S}$, we are
interested in the \emph{global} utility $f\left(\boldsymbol{a},\mathbf{S}\right)$
of the network, which is defined as the sum of the \emph{local} utility
function $u_{i}\left(\boldsymbol{a},\mathbf{S}\right)$ of each node~$i$,
\emph{i.e.}, 
\[
f\left(\boldsymbol{a},\mathbf{S}\right)=\sum_{i\in\mathcal{N}}u_{i}\left(\boldsymbol{a},\mathbf{S}\right).
\]
The network performance is then characterized by the\emph{ average
global} utility 
\[
F\left(\boldsymbol{a}\right)=\mathbb{E}_{\mathbf{S}}\left(f\left(\boldsymbol{a},\mathbf{S}\right)\right).
\]

In this work, we consider a challenging setting that nodes do not
have the knowledge of $\mathbf{S}_{k}$ nor the closed-form expression
of the utility functions. Each node~$i$ only has a numerical estimation
$\widetilde{u}_{i,k}$ of its local utility $u_{i}\left(\boldsymbol{a}_{k},\mathbf{S}_{k}\right)$
at each timeslot. Assume that
\begin{equation}
\widetilde{u}_{i,k}=u_{i}\left(\boldsymbol{a}_{k},\mathbf{S}_{k}\right)+\eta_{i,k},\label{eq:u_noisy}
\end{equation}
with $\eta_{i,k}$ some additive noise. Nodes can communicate with
each other so that each node~$i$ can get a approximate value $\widetilde{f}_{i,k}$
of the global utility $f_{i}\left(\boldsymbol{a}_{k},\mathbf{S}_{k}\right)$. 

As a summary, our aim is to propose some distributed solution of the
following problem 
\begin{equation}
\begin{cases}
\textrm{maximize} & F\left(\boldsymbol{a}\right)=\mathbb{E}_{\mathbf{S}}\left(f\left(\boldsymbol{a},\mathbf{S}\right)\right),\\
\textrm{subject to} & a_{i}\in\mathcal{A}_{i},\:\forall i\in\mathcal{N}.
\end{cases}\label{eq:a_opt_def}
\end{equation}
An application example is introduced in Section~\ref{sec:Motivating-Example}
to highlight the interest of this problem. 

\subsection{Basic assumptions}

We present the basic assumptions considered in this paper in order
to guarantee the performance of our proposed algorithm. 

Denote $\boldsymbol{a}^{*}$ as the solution of the problem (\ref{eq:a_opt_def}).
Since the existence of $\boldsymbol{a}^{*}$ can be ensured by the
concavity of the objective function $F\left(\boldsymbol{a}\right)$,
we have the following assumption.
\begin{itemize}
\item \emph{A1}. (properties of objective function) Both $F_{i}'\left(\boldsymbol{a}\right)$
and $F_{i,j}''\left(\boldsymbol{a}\right)$ exist continuously. There
exists $\boldsymbol{a}^{*}\in\mathcal{A}$ such that $F_{i}'\left(\boldsymbol{a}^{*}\right)=0$
and $F_{i,i}''\left(\boldsymbol{a}^{*}\right)<0$, $\forall i\in\mathcal{N}$.
The objective function is strictly concave, \emph{i.e.},
\begin{equation}
\left(\boldsymbol{a}-\boldsymbol{a}^{*}\right)^{T}\cdot\nabla F\left(\boldsymbol{a}\right)\leq0,\:\forall\boldsymbol{a}\in\mathcal{A}.\label{eq:concave}
\end{equation}
 Besides, for any $i\in\mathcal{N}$ and $j\in\mathcal{N}$, there
exists a constant $\alpha_{1}\in\mathbb{R}^{+}$ such that 
\begin{equation}
\left|F_{i,j}''\left(\boldsymbol{a}\right)\right|\leq\alpha_{1}.\label{eq:assumption_F}
\end{equation}
\end{itemize}
We have some further assumptions on the local utility functions, which
will be useful in our analysis.
\begin{itemize}
\item \emph{A2}. (properties of local utility function) For any $i\in\mathcal{N}$,
the function $\boldsymbol{a}\longmapsto u_{i}\left(\boldsymbol{a},\mathbf{S}\right)$
is Lipschitz with Lipschitz constant $L_{\mathbf{S}}$, \emph{i.e.},
\begin{equation}
\left\Vert u_{i}\left(\boldsymbol{a},\mathbf{S}\right)-u_{i}\left(\widetilde{\boldsymbol{a}},\mathbf{S}\right)\right\Vert \leq L_{\mathbf{S}}\left\Vert \boldsymbol{a}-\widetilde{\boldsymbol{a}}\right\Vert .\label{eq:lipschitz_u}
\end{equation}
Besides, $\mathbb{E}_{\mathbf{S}}\left(u_{i}\left(\boldsymbol{a},\mathbf{S}\right)\right)<\infty$,
$\forall i\in\mathcal{N}$, so that $F\left(\boldsymbol{a}\right)$
is also bounded.
\end{itemize}
From \emph{A2}, we can easily deduce that $\boldsymbol{a}\longmapsto f\left(\boldsymbol{a},\mathbf{S}\right)$
is also Lipschitz with Lipschitz constant $NL_{\mathbf{S}}$, since
\begin{align}
 & \left\Vert f\left(\boldsymbol{a},\mathbf{S}\right)-f\left(\widetilde{\boldsymbol{a}},\mathbf{S}\right)\right\Vert =\left\Vert \sum_{i\in\mathcal{N}}\left(u_{i}\left(\boldsymbol{a},\mathbf{S}\right)-u_{i}\left(\widetilde{\boldsymbol{a}},\mathbf{S}\right)\right)\right\Vert \nonumber \\
 & \leq\sum_{i\in\mathcal{N}}\left\Vert u_{i}\left(\boldsymbol{a},\mathbf{S}\right)-u_{i}\left(\widetilde{\boldsymbol{a}},\mathbf{S}\right)\right\Vert \leq NL_{\mathbf{S}}\left\Vert \boldsymbol{a}-\widetilde{\boldsymbol{a}}\right\Vert .\label{eq:Lipschitz_f}
\end{align}

In the end, we consider a common assumption on the noise term $\eta_{i,k}$
introduced in (\ref{eq:u_noisy}).
\begin{itemize}
\item \emph{A3}. (properties of additive noise) The noise $\eta_{i,k}$
is zero-mean, uncorrelated, and has bounded variance, \emph{i.e.},
for any $i\in\mathcal{N}$, we have $\mathbb{E}\left(\eta_{i,k}\right)=0$,
$\mathbb{E}\left(\eta_{i,k}^{2}\right)=\alpha_{4}<\infty$, and $\mathbb{E}\left(\eta_{i,k}\eta_{j,k}\right)=0$
if $i\neq j$.
\end{itemize}

\section{Motivating examples\label{sec:Motivating-Example}}

\begin{figure}
\begin{centering}
\includegraphics[width=1\columnwidth]{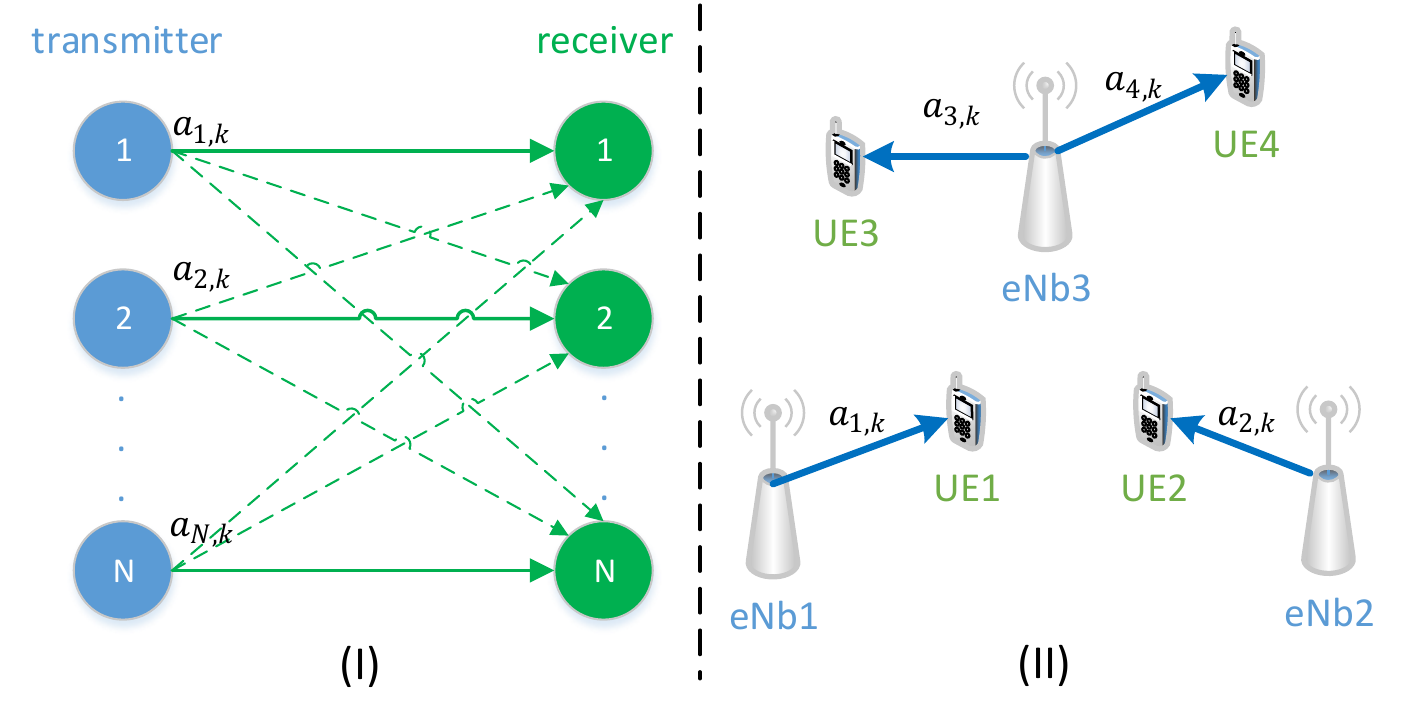}
\par\end{centering}
\caption{(I) A D2D network with $N$ transmitter-receiver pairs; (II) Downlink
multi-cell network, each eNB (or base station) may serve multiple
UEs (users). \label{fig:A-wireless-network}}
\end{figure}

This section provides some examples and shows the limit of the classical
gradient-based methods.

We consider the power allocation problem in a general network with
$N$ transmitter-receiver links. A link here can be seen as a node
in our system model as presented in Section~\ref{sec:System-Model}.
For example,
\begin{itemize}
\item in a D2D network with $N$ transmitter-receiver pairs, each transmitter
communicates with its associated receiver and different links interfere
among each other, see Figure~\ref{fig:A-wireless-network}~(I).
\item in a multi-cell network, each base station may serve multiple users.
We focus on the downlink, \emph{i.e.}, a user is seen as its receiver
and its associated base station is seen as a transmitter, see Figure~\ref{fig:A-wireless-network}~(II).
\end{itemize}
In both models, the action $a_{i,k}$ is in fact the transmission
power of transmitter~$i$ at timeslot~$k$, of which the value cannot
exceed the maximum transmission power $a_{\textrm{max}}$. Thus we
have $\mathcal{A}_{i}=\left[0,a_{\textrm{max}}\right]$, $\forall i\in\mathcal{N}$.
The environment state matrix $\mathbf{S}_{k}$ represents the time-varying
channel state of the network. More precisely, $\mathbf{S}_{k}=\left[s_{ij,k}\right]_{i\in\mathcal{N},j\in\mathcal{N}}$,
in which each element $s_{ij,k}$ denotes the the channel gain between
transmitter $i$ and receiver $j$ at timeslot $k$. 

The utility function can be various depending on different applications,
\emph{e.g.}, throughput and energy efficiency \cite{Mertikopoulo16}.

In our first example, the local utility of each node~$i$ is given
by
\begin{equation}
u_{i}\left(\boldsymbol{a}_{k},\mathbf{S}_{k}\right)=\omega\log\left(1+r_{i,k}\right)-\kappa a_{i,k},\label{eq:u_function-1}
\end{equation}
where $\omega,\kappa\in\mathbb{R}^{+}$, $\kappa a_{i,k}$ represents
the energy cost of transmission, and $r_{i,k}$ denotes the bit rate
given by 
\begin{equation}
r_{i,k}=\log\left(1+\mathtt{SINR}_{i,k}\right)\textrm{ with }\mathtt{SINR}_{i,k}=\frac{a_{i,k}s_{ii,k}}{\sigma^{2}+\sum_{j\neq i}a_{j,k}s_{ij,k}}.\label{eq:rate_def}
\end{equation}
Note that the maximization of $\log$-function of the bit rate is
of type \emph{proportional fairness}, which is used to ensure fairness
among the nodes in the network. With the above notations, the global
utility function can be written as
\begin{align}
 & f^{\left(1\right)}\left(\boldsymbol{a}_{k},\mathbf{S}_{k}\right)=-\kappa\sum_{i\in\mathcal{N}}a_{i,k}\nonumber \\
 & +\omega\sum_{i\in\mathcal{N}}\log\left(1+\log\left(1+\frac{a_{i,k}s_{ii,k}}{\sigma^{2}+\sum_{j\neq i}a_{j,k}s_{ji,k}}\right)\right).\label{eq:u_function}
\end{align}
It is straightforward to show that $f^{\left(1\right)}\left(\boldsymbol{a}_{k},\mathbf{S}_{k}\right)$
is concave with respect to $\boldsymbol{a}_{k}$, hence the existence
of the optimum $\boldsymbol{a}^{*}$ can be guaranteed. 

In order to maximize the average global utility $\mathbb{E}_{\mathbf{S}_{k}}\left(f^{\left(1\right)}\left(\boldsymbol{a}_{k},\mathbf{S}_{k}\right)\right)$,
one may apply the classical stochastic approximation method described
by (\ref{eq:algo_update}). An essential step is that, each transmitter
or receiver should be able to calculate the partial derivative, \emph{i.e.},
\begin{align}
 & \widehat{g}_{i,k}=\frac{\partial f^{\left(1\right)}}{\partial a_{i,k}}=\frac{\omega\mathtt{SINR}_{i,k}}{\left(1+r_{i,k}\right)a_{i,k}}-\kappa\nonumber \\
 & \quad-\sum_{n\in\mathcal{N}}\frac{\omega\mathtt{SINR}_{n,k}^{2}}{\left(1+r_{n,k}\right)\left(1+\mathtt{SINR}_{n,k}\right)}\frac{s_{ni,k}}{a_{n,k}s_{nn,k}}.\label{eq:deri_theo}
\end{align}
From (\ref{eq:deri_theo}), we find that the direct calculation of
the partial derivative is complicated and nodes should exchange much
information: Each node~$i$ should know the values of $\mathtt{SINR}_{n,k}$,
the cross-channel gain $s_{ni,k}$, as well as $a_{n,k}s_{nn,k}$
of all nodes $n\in\mathcal{N}$. Moreover, in the situation where
the channel is time-varying, it is not realistic for any receiver~$n$
to estimate the direct-channel gain $s_{nn,k}$ and all the cross-channel
gain $s_{nj,k}$, $\forall j\in\mathcal{N}\setminus\left\{ n\right\} $. 

We consider the sum-rate maximization problem as a second example,
\emph{i.e.}, to maximize $y_{k}=\sum_{i\in\mathcal{N}}r_{i,k}$. The
challenge come from the fact that $y_{k}$ is not concave with respect
to $a_{i,k}$ if the rate $r_{i,k}$ is given by (\ref{eq:rate_def}).
For this reason, we have to consider the approximation of $r_{i,k}$
and some variable change to make the objective function concave, which
is a well known problem \cite{tan2013fast}. It is common to use change
of variable (\emph{i.e.}, consider $\textrm{e}^{a_{i,k}}$ as the
transmission power instead of $a_{i,k}$) and consider the approximation
$r_{i,k}\approx\log\left(\mathtt{SINR}_{i,k}\right)$ \cite{tan2013fast},
so that the global utility function is written as 
\begin{align}
 & f^{\left(2\right)}\!\left(\boldsymbol{a}_{k},\mathbf{S}_{k}\right)\!=\!\omega\!\sum_{i\in\mathcal{N}}\!\log\!\left(\!\!\frac{s_{ii,k}\textrm{e}^{a_{i,k}}}{\sigma^{2}+\sum_{j\neq i}s_{ji,k}\textrm{e}^{a_{j,k}}}\!\!\right)\!-\kappa\!\sum_{i\in\mathcal{N}}\!\textrm{e}^{a_{i,k}}.\label{eq:u_function-1-1}
\end{align}
It is straightforward to show the concavity of $f^{\left(2\right)}$.
Similar to (\ref{eq:deri_theo}), we evaluate
\begin{align}
\widehat{g}_{i,k} & =\frac{\partial f^{\left(2\right)}}{\partial a_{i,k}}\nonumber \\
 & =\omega-\omega\!\sum_{n\in\mathcal{N}}\!\!\frac{s_{in,k}\textrm{e}^{a_{i,k}}}{\sigma^{2}+\sum_{j\neq n}\delta_{j,k}s_{jn,k}\textrm{e}^{a_{j,k}}}\!-\kappa\!\sum_{i\in\mathcal{N}}\!\textrm{e}^{a_{i,k}},\label{eq:deri_theo-1}
\end{align}
of which the calculation also requires much information, such as the
cross-channel gain $s_{in,k}$ $\forall n\in\mathcal{N}\setminus\left\{ i\right\} $,
and all the interference estimated by each receiver. All the channel
information has to be estimated and exchanged by each active node,
which is a huge burden for the network. 

The two examples clearly shows the limit of the classical methods
and motivates us to propose some novel solution where low informational
exchange are required. In this paper, we consider that the nodes can
only exchange their numerical approximation of local utilities. It
is worth mentioning that the information exchange in our setting is
much less than the gradient based method. We present our distributed
optimization algorithms as well as their performance, in the situations
where each node has the \textit{complete} or \textit{incomplete} knowledge
of the local utilities of the other nodes, respectively. It is worth
mentioning that, apart from the examples presented in this section,
the solution proposed in this paper can also be applied to other type
of problems such beamforming, coordinated multipoint (CoMP) \cite{irmer2011coordinated},
and so on.

\section{Distributed optimization algorithm using stochastic perturbation\label{sec:Distributed-allocation-algorithm}}

This section presents a first version of our distributed optimization
algorithm. We assume that each node is \emph{always} able to collect
the numerical estimation of local utilities from all the other nodes.
In this situation, each node can evaluate the numerical value of the
global utility function at each iteration (or timeslot)~$k$ by applying
\begin{align}
\widetilde{f}\left(\boldsymbol{a},\mathbf{S}_{k}\right) & =\sum_{i\in\mathcal{N}}\widetilde{u}_{i,k}=\sum_{i\in\mathcal{N}}\left(u_{i}\left(\boldsymbol{a},\mathbf{S}_{k}\right)+\eta_{i,k}\right)\nonumber \\
 & =f\left(\boldsymbol{a},\mathbf{S}_{k}\right)+\sum_{i\in\mathcal{N}}\eta_{i,k},\label{eq:f_approx}
\end{align}
since each node knows the \emph{complete} information of $\widetilde{u}_{i,k}$
for any $i\in\mathcal{N}$. We first consider an unconstrained optimization
problem, \emph{i.e.}, $\mathcal{A}=\mathbb{R}^{N}$, in Section~\ref{subsec:Algorithm_unc}.
The constrained optimization problem with $\mathcal{A}_{i}=\left[a_{i,\min},a_{i,\max}\right]$,
$\forall i\in\mathcal{N}$ is then presented in Section~\ref{sec:constrained}.

\subsection{Algorithm\label{subsec:Algorithm_unc}}

The distributed optimization algorithm using stochastic perturbation
(DOSP) is presented in Algorithm~\ref{alg:ESSP-based-Algorithm}. 

\begin{algorithm}[h]
\caption{\label{alg:ESSP-based-Algorithm}DOSP Algorithm for each node~$i$}
\begin{enumerate}
\item Initialize $k=0$ and set the action $a_{i,0}$ randomly.
\item Generate a random variable $\Phi_{i,k}$, perform action $a_{i,k}+\gamma_{k}\Phi_{i,k}$.
\item Estimate $\widetilde{u}_{i,k}$, exchange its value with the other
nodes and calculate $\widetilde{f}\left(\boldsymbol{a}_{k}+\gamma_{k}\boldsymbol{\Phi}_{k},\mathbf{S}_{k}\right)=\sum_{j\in\mathcal{N}}\widetilde{u}_{i,k}.$
\item Update $a_{i,k+1}$ according to equation (\ref{eq:algo_sp}).
\item $k=k+1$, go to 2.
\end{enumerate}
\end{algorithm}

At each iteration~$k$, an arbitrary reference node~$i$ updates
its action by applying 
\begin{equation}
a_{i,k+1}=a_{i,k}+\beta_{k}\Phi_{i,k}\widetilde{f}\left(\boldsymbol{a}_{k}+\gamma_{k}\boldsymbol{\Phi}_{k},\mathbf{S}_{k}\right),\label{eq:algo_sp}
\end{equation}
in which $\beta_{k}$ and $\gamma_{k}$ are vanishing step-sizes,
$\Phi_{i,k}$ is randomly generated by each node~$i$ and $\boldsymbol{\Phi}_{k}=\left[\Phi_{1,k},\ldots,\Phi_{N,k}\right]$.
Recall that the approximation $\widetilde{f}$ of the global utility
is calculated by each node using (\ref{eq:f_approx}), of which the
value depends on the actual action performed by each node $\widehat{a}_{i,k}=a_{i,k}+\gamma_{k}\Phi_{i,k}$
and the environment state matrix $\mathbf{S}_{k}$. Note that $\widehat{a}_{i,k}$
is very close to $a_{i,k}$ when $k$ is large as $\gamma_{k}$ is
vanishing. An example is presented in Section~\ref{subsec:Application-example}
to describe in detail the application of Algorithm~\ref{alg:ESSP-based-Algorithm}
in practice.

Obviously, (\ref{eq:algo_sp}) can be written in the general form
(\ref{eq:algo_update}), in which 
\begin{equation}
\widehat{g}_{i,k}=\Phi_{i,k}\widetilde{f}\left(\boldsymbol{a}_{k}+\gamma_{k}\boldsymbol{\Phi}_{k},\mathbf{S}_{k}\right).\label{eq:g_est}
\end{equation}
As we will discuss in Section~\ref{subsec:Convergence-results},
$\widehat{g}_{i,k}$ can be an asymptotically unbiased estimation
of the partial derivative $\partial F/\partial a_{i}$ and $\boldsymbol{a}_{k}$
can converge to $\boldsymbol{a}^{*}$, under the condition that the
parameters $\beta_{k}$, $\gamma_{k}$, and $\boldsymbol{\Phi}_{k}$
are properly chosen. We state in what follows the desirable properties
of these parameters.
\begin{itemize}
\item \emph{A4}. (properties of step-sizes) Both $\beta_{k}$ and $\gamma_{k}$
take real positive values with $\lim_{k\rightarrow\infty}\beta_{k}=\lim_{k\rightarrow\infty}\gamma_{k}=0$,
besides,
\[
\sum_{k=1}^{\infty}\beta_{k}\gamma_{k}=\infty,\qquad\sum_{k=1}^{\infty}\beta_{k}^{2}<\infty.
\]
\item \emph{A5}. (properties of random perturbation) The elements of $\boldsymbol{\Phi}_{k}$
are i.i.d. with $\mathbb{E}\left(\Phi_{i,k}\Phi_{j,k}\right)=0$,
$\forall i\neq j$. There exist $\alpha_{2}>0$ and $\alpha_{3}>0$
such that
\[
\mathbb{E}\left(\Phi_{i,k}^{2}\right)=\alpha_{2},\quad\left|\Phi_{i,k}\right|\leq\alpha_{3}.
\]
\end{itemize}
The conditions on the parameters can be easily achieved. We show in
Example~\ref{rem:condition} a common setting of these parameters,
which are also used to obtain the simulation results to be presented
in Section~\ref{sec:Simulation-Results}.

\begin{eg}\label{rem:condition}An easiest choice of the probability
distribution of $\Phi_{i,k}$ is the symmetrical Bernoulli distribution
with $\Phi_{i,k}\in\left\{ -1,1\right\} $ and $\mathbb{P}\left(\Phi_{i,k}=1\right)=\mathbb{P}\left(\Phi_{i,k}=-1\right)=0.5$,
$\forall i,k$. We can verify the conditions in \emph{A5} with $\alpha_{2}=\alpha_{3}=1$.

Let $\beta_{k}=\beta_{0}\left(k+1\right)^{-\nu_{1}}$ and $\gamma_{k}=\gamma_{0}\left(k+1\right)^{-\nu_{2}}$
with the constants $\beta_{0},\gamma_{0},\nu_{1},\nu_{2}\in\mathbb{R}^{+}$,
so that both $\beta_{k}$ and $\gamma_{k}$ are vanishing. Since $\sum_{k=1}^{\infty}\beta_{k}^{2}$
converges if $\nu_{1}>0.5$; $\sum_{k=1}^{\infty}\beta_{k}\gamma_{k}$
diverges if $\nu_{1}+\nu_{2}\leq1$. Clearly, there exist pairs of
$\nu_{1}$ and $\nu_{2}$ to make $\beta_{k}$ and $\gamma_{k}$ satisfy
the conditions in \emph{A4}. \end{eg}

\begin{remark}\label{rem:algo_diff}The proposed algorithm has similar
shape compared with the other existed methods proposed in \cite{manzie2009extremum,frihauf2012nash,hanif2012convergence}.
The difference between our solution and the sine perturbation based
method \cite{frihauf2012nash,hanif2012convergence} is that, we use
a random vector $\boldsymbol{\Phi}_{k}$ instead of some deterministic
sine functions as the perturbation term. Comparing (\ref{eq:algo_update-1})
and (\ref{eq:algo_sp}), we can see that the amplitude of random perturbation
is vanishing in our algorithm, which is not the case in the algorithm
presented in \cite{manzie2009extremum}. \end{remark}

\subsection{Application example\label{subsec:Application-example}}

This section presents the application of Algorithm~\ref{alg:ESSP-based-Algorithm}
to perform resource allocation in a D2D network, in order to highlight
the interest of our solution. We focus on the requirement arisen by
Algorithm~\ref{alg:ESSP-based-Algorithm}, in terms of computation,
memory and informational exchange. 

\begin{figure}
\begin{centering}
\includegraphics[width=1\columnwidth]{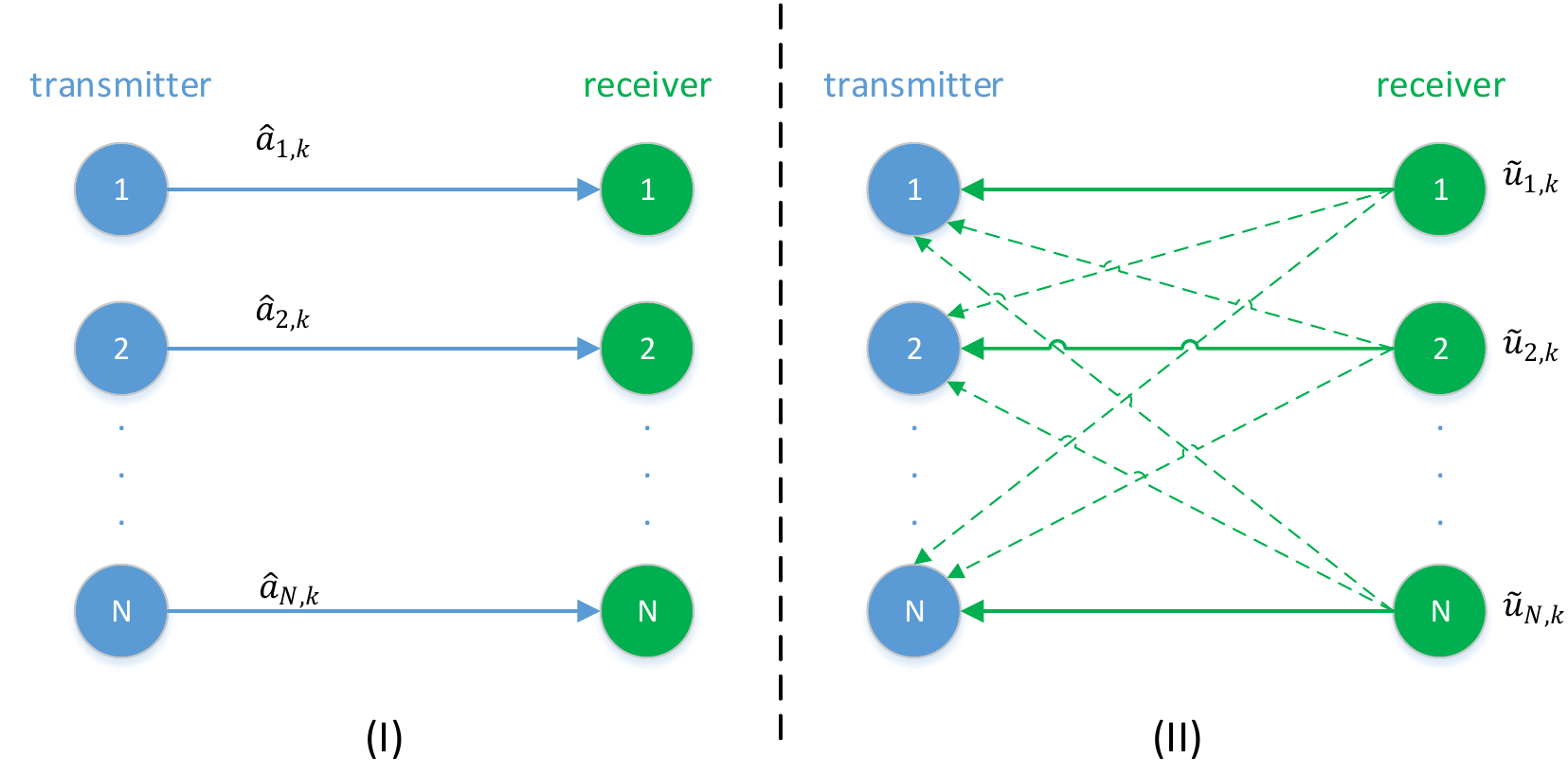}
\par\end{centering}
\caption{\label{fig:example_ex}(I) At iteration $k$, each transmitter $i$
transmits to its associated receiver with transmission power $\widehat{a}_{i,k}$;
(II) Each receiver $i$ sends the approximate local utility $\widetilde{u}_{i,k}$
to its transmitter, every transmitter can receive and decode this
feedback information.}
\end{figure}

Figure~\ref{fig:example_ex} briefly shows the algorithm procedure
during one iteration. Recall that $\widehat{a}_{i,k}$ denotes the
actual value of the action set by transmitter~$i$ at iteration~$k$.
In order to update $\widehat{a}_{i,k}$, each \emph{transmitter}~$i$
needs to
\begin{itemize}
\item independently and randomly generate a scalar $\Phi_{i,k}$ under the
condition \emph{A5;}
\item use a pre-defined vanishing sequence $\gamma_{k}$ which is common
for each link;
\item independently update $a_{i,k}$ by applying (\ref{eq:algo_sp}), more
details will be provided soon.
\end{itemize}
Then $\widehat{a}_{i,k}$ is given by $\widehat{a}_{i,k}=a_{i,k}+\gamma_{k}\Phi_{i,k}$. 

Each transmitter~$i$ transmits to its associated receiver with the
transmission power of value $\widehat{a}_{i,k}$. The associated receiver~$i$
should be able to approximate the numerical value of its local utility
$\widetilde{u}_{i,k}$ and send this value to transmitter~$i$ as
a feedback. All the transmitters can evaluate $\widetilde{f}\left(\widehat{\boldsymbol{a}}_{k}\right)$
using (\ref{eq:f_approx}) under the assumption that every transmitter
is able to receive and decode the feedback from all the receivers.

Then iteration~$k+1$ starts, each transmitter~$i$ needs to update
its power allocation strategy, what is required is listed as follows:
\begin{itemize}
\item use a pre-defined vanishing sequence $\beta_{k}$;
\item reuse of the local random value $\Phi_{i,k}$, which was generated
by transmitter~$i$ at iteration~$k$. It means that the value of
$\Phi_{i,k}$ should be saved temporarily
\item use the numerical value of $\widetilde{f}_{k}\left(\widehat{\boldsymbol{a}}_{k}\right)$,
which has been explained already.
\end{itemize}
Each transmitter can then update $a_{i,k+1}$ independently using
(\ref{eq:algo_sp}). 

In the following step, each transmitter updates $\widehat{a}_{i,k+1}$
in the same way as the previous iteration, \emph{i.e.}, $\widehat{a}_{i,k+1}=a_{i,k+1}+\gamma_{k+1}\Phi_{i,k+1},$
with $\Phi_{i,k+1}$ a newly generated pseudo-random value. 

We can see that Algorithm~\ref{alg:ESSP-based-Algorithm} can be
easily applied in a network of multiple links: the algorithm itself
has low complexity and each receiver only needs to feedback one quantity
($\widetilde{u}_{i,k}$) per iteration to perform the algorithm. 

\subsection{Convergence results\label{subsec:Convergence-results}}

This section investigates the asymptotic behavior of Algorithm~\ref{alg:ESSP-based-Algorithm}.
For any integer $k\geq0$, we consider the divergence 
\begin{equation}
d_{k}=\left\Vert \boldsymbol{a}_{k}-\boldsymbol{a}^{*}\right\Vert ^{2}\label{eq:divergence_def}
\end{equation}
to describe the difference between the actual action $\boldsymbol{a}_{k}$
and the optimal action $\boldsymbol{a}^{*}$. Our aim is to show that
$d_{k}\rightarrow0$ almost surely as $k\rightarrow\infty$.

In order to explain the reason behind the update rule (\ref{eq:algo_sp}),
we rewrite it in the generalized Robbins-Monro form \cite{kushner2012stochastic},
\emph{i.e.}, 
\begin{align}
 & \boldsymbol{a}_{k+1}=\boldsymbol{a}_{k}+\beta_{k}\widehat{\boldsymbol{g}}_{k}\nonumber \\
 & =\boldsymbol{a}_{k}+\beta_{k}\left(\alpha_{2}\gamma_{k}\nabla F\left(\boldsymbol{a}_{k}\right)+\overline{\boldsymbol{g}}_{k}-\alpha_{2}\gamma_{k}\nabla F\left(\boldsymbol{a}_{k}\right)+\widehat{\boldsymbol{g}}_{k}-\overline{\boldsymbol{g}}_{k}\right)\nonumber \\
 & =\boldsymbol{a}_{k}+\alpha_{2}\beta_{k}\gamma_{k}\left(\nabla F\left(\boldsymbol{a}_{k}\right)+\boldsymbol{b}_{k}+\frac{\boldsymbol{e}_{k}}{\alpha_{2}\gamma_{k}}\right),\label{eq:RM_form}
\end{align}
where $\overline{\boldsymbol{g}}_{k}$ represents the expected value
of $\widehat{\boldsymbol{g}}_{k}$ with respect to all the stochastic
terms (including $\boldsymbol{\Phi}$, $\mathbf{S}$, and $\boldsymbol{\eta}$)
for a given $\boldsymbol{a}_{k}$, \emph{i.e.},
\begin{equation}
\overline{\boldsymbol{g}}_{k}=\mathbb{E}_{\mathbf{S},\boldsymbol{\Phi},\boldsymbol{\eta}}\left(\widehat{\boldsymbol{g}}_{k}\right),\label{eq:g_aver}
\end{equation}
note that we prefer to highlight the stochastic terms in (\ref{eq:g_aver}),
an alternative way is to write (\ref{eq:g_aver}) as $\overline{\boldsymbol{g}}_{k}=\mathbb{E}\left(\widehat{\boldsymbol{g}}_{k}\mid\boldsymbol{a}_{k}\right)$;
$\boldsymbol{b}_{k}$ denotes the estimation bias between $\overline{\boldsymbol{g}}_{k}$
and the actual gradient of the average objective function $F$, \emph{i.e.},
\begin{align}
\boldsymbol{b}_{k} & =\frac{\overline{\boldsymbol{g}}_{k}}{\alpha_{2}\gamma_{k}}-\nabla F\left(\boldsymbol{a}_{k}\right);\label{eq:bias_def}
\end{align}
and $\boldsymbol{e}_{k}$ can be seen as the stochastic noise, which
is the difference between the value of a single realization of $\widehat{\boldsymbol{g}}_{k}$
and its average $\overline{\boldsymbol{g}}_{k}$ as defined in (\ref{eq:g_aver}),
\emph{i.e.}, 
\begin{equation}
\boldsymbol{e}_{k}=\widehat{\boldsymbol{g}}_{k}-\overline{\boldsymbol{g}}_{k}.\label{eq:sto_n_def}
\end{equation}

\begin{remark}The analysis presented in this work is challenging
and different from the existed results. An explicit difference comes
from the unique feature of the algorithm itself as discussed in Remark~\ref{rem:algo_diff}:
we are using a different method to estimate the gradient. Besides,
the objective function is stochastic with general form in our problem,
while it is considered as static in \cite{spall1992multivariate,spall1997one}
and it is assumed to be static and quadratic in \cite{manzie2009extremum}.\end{remark}

To perform the analysis of convergence, we have to investigate the
properties of $\boldsymbol{b}_{k}$ and $\boldsymbol{e}_{k}$. 
\begin{lem}
\label{lem:bias}If all the conditions in A1-A5 are satisfied, then
\begin{equation}
\left\Vert \boldsymbol{b}_{k}\right\Vert \leq\gamma_{k}N^{\frac{5}{2}}\frac{\alpha_{3}^{3}\alpha_{1}}{2\alpha_{2}}=O\left(\gamma_{k}\right),\label{eq:bias_bound}
\end{equation}
which implies that $\left\Vert \boldsymbol{b}_{k}\right\Vert \rightarrow0$
as $k\rightarrow\infty$.
\end{lem}
\begin{IEEEproof}
See Appendix \ref{subsec:Proof-bias}.
\end{IEEEproof}
Lemma~\ref{lem:bias} implies that $\widehat{\boldsymbol{g}}_{k}$
defined in Algorithm~\ref{alg:ESSP-based-Algorithm} is a reasonable
estimator of the gradient with vanishing bias. 
\begin{lem}
\label{lem:stochastic_noise}If all the conditions in A1-A5 are satisfied
and $\left\Vert \boldsymbol{a}_{k}\right\Vert <\infty$ almost surely,
then for any constant $\rho>0$, we have 
\begin{equation}
\lim_{K\rightarrow\infty}\mathbb{P}\left(\sup_{K'\geq K}\left\Vert \sum_{k=K}^{K'}\beta_{k}\boldsymbol{e}_{k}\right\Vert \geq\rho\right)=0,\quad\forall\rho>0.\label{eq:pb_toprove}
\end{equation}
\end{lem}
\begin{IEEEproof}
See Appendix \ref{subsec:Proof-e}.
\end{IEEEproof}
Based on the results in Lemma~\ref{lem:bias} and in Lemma~\ref{lem:stochastic_noise},
we can build conditions under which $\boldsymbol{a}_{k}\rightarrow\boldsymbol{a}^{*}$
almost surely.

\begin{mythm}\label{prop:converge_rp}If all the conditions in A1-A5
are satisfied and $\left\Vert \boldsymbol{a}_{k}\right\Vert <\infty$
almost surely, then $\boldsymbol{a}_{k}\rightarrow\boldsymbol{a}^{*}$
as $k\rightarrow\infty$ almost surely by applying Algorithm~\ref{alg:ESSP-based-Algorithm}.
\end{mythm}
\begin{IEEEproof}
See Appendix \ref{subsec:proof_conv}.
\end{IEEEproof}

\subsection{Distributed optimization under constraints\label{sec:constrained}}

In this section, we consider the constrained optimization problem,
in which the action of each node takes value from an interval, \emph{i.e.},
$\mathcal{A}_{i}=\left[a_{i,\min},a_{i,\max}\right]$, $\forall i\in\mathcal{N}$.
We assume that $\boldsymbol{a}^{*}$ is not on the boundary of the
feasible set $\mathcal{A}$, \emph{i.e.}, $a_{i}^{*}\in\left(a_{i,\min},a_{i,\max}\right)$,
$\forall i\in\mathcal{N}$. For example, in the power allocation problem,
$a_{i,k}$ presents the transmission power of transmitter~$i$, which
should be positive and not larger than a maximum value. 

Recall that the actually performed action by nodes is $\widehat{\boldsymbol{a}}_{k}=\boldsymbol{a}_{k}+\gamma_{k}\boldsymbol{\Phi}_{k}$.
At each iteration, we need to ensure that $\widehat{a}_{i,k}\in\left[a_{i,\min},a_{i,\max}\right]$.
A direct solution is to introduce a projection to the algorithm, \emph{i.e.},
(\ref{eq:algo_sp}) turns to
\begin{equation}
\boldsymbol{a}_{k+1}=\mathtt{Proj}\left(\boldsymbol{a}_{k}+\beta_{k}\widehat{\boldsymbol{g}}_{k}\right),\label{eq:proj_algo1}
\end{equation}
in which for any $i\in\mathcal{N}$,
\begin{align}
a_{i,k+1} & =\min\left\{ \max\left\{ a_{i,k}+\beta_{k}\widehat{g}_{i,k},a_{i,\min}+\alpha_{3}\gamma_{k+1}\right\} ,\right.\nonumber \\
 & \qquad\qquad\qquad\left.a_{i,\max}-\alpha_{3}\gamma_{k+1}\right\} .\label{eq:proj_sp2}
\end{align}
As $\left|\Phi_{i,k}\right|\leq\alpha_{3}$ by \emph{A5}, (\ref{eq:proj_sp2})
leads to $a_{i,k+1}+\gamma_{k+1}\left|\Phi_{i,k+1}\right|\in\left[a_{i,\min},a_{i,\max}\right]$. 

\begin{remark}\label{rem_const}In order to write (\ref{eq:proj_sp2})
into the form similar to (\ref{eq:RM_form}), one has to re-define
the bias term $b_{i,k}$ and the stochastic noise $e_{i,k}$ because
of the operator $\mathtt{Proj}$. Therefore, it is not straightforward
to deduce of the convergence of the constrained optimization algorithm
from the results presented in Section~\ref{subsec:Convergence-results}.\end{remark}

\begin{mythm}\label{prop:converge_constraint}In the constrained
optimization problem where $\mathcal{A}_{i}=\left[a_{i,\min},a_{i,\max}\right]$
and $a_{i}^{*}\in\left(a_{i,\min},a_{i,\max}\right)$, $\forall i\in\mathcal{N}$,
by applying the projection~(\ref{eq:proj_algo1}), we have $\boldsymbol{a}_{k}\rightarrow\boldsymbol{a}^{*}$
as $k\rightarrow\infty$ almost surely under the assumptions A1-A5.
\end{mythm}
\begin{IEEEproof}
See Appendix~\ref{subsec:Proof-constrained}.
\end{IEEEproof}

\section{Optimization Algorithm with incomplete information of utilities of
other nodes \label{sec:ESSP-Algorithm-incomplete}}

A limit of Algorithm~\ref{alg:ESSP-based-Algorithm} is that each
node is required to know the local utility of all the other nodes.
Such issue is significant as there are many nodes in the network.
It is thus important to consider a more realistic situation where
a node only has has the knowledge of the local utilities of a subset
$\mathcal{I}_{i,k}$ of nodes, with $\mathcal{I}_{i,k}\subseteq\mathcal{N}\setminus\left\{ i\right\} $.
Throughout this section, we have the following assumption:
\begin{itemize}
\item \emph{A6}). at any iteration~$k$, an arbitrary node~$i$ knows
the utility $\widetilde{u}_{j,k}$ of another node~$j$ with a constant
probability $p\in\left(0,1\right]$, $i.e.$, the elements contained
in the set $\mathcal{I}_{i,k}$ is random, for any $j\neq i$, we
have
\begin{equation}
\mathbb{P}\left(j\in\mathcal{I}_{i,k}\right)=p,\qquad\mathbb{P}\left(j\notin\mathcal{I}_{i,k}\right)=1-p.\label{eq:p}
\end{equation}
\end{itemize}
Notice that we do not assume any specified network topology and each
node $i$ has a different and \textit{independent} set $\mathcal{I}_{i,k}$.

We propose a modified algorithm and then show its asymptotic performance.
The algorithm is described in Algorithm~\ref{alg:ESSP-based-Algorithm-1}.
The main difference between Algorithm~\ref{alg:ESSP-based-Algorithm}
and Algorithm~\ref{alg:ESSP-based-Algorithm-1} comes from the approximation
of the objective function, \emph{i.e.},
\begin{align}
 & \widetilde{f}_{i}^{(\textrm{I})}\left(\boldsymbol{a},\mathbf{S}_{k},\mathcal{I}_{i,k}\right)\nonumber \\
 & =\begin{cases}
\widetilde{u}_{i,k}+\frac{N-1}{\left|\mathcal{I}_{i,k}\right|}\sum_{j\in\mathcal{I}_{i,k}}\widetilde{u}_{j,k}, & \textrm{if }\left|\mathcal{I}_{i,k}\right|\neq0,\\
0, & \textrm{if }\left|\mathcal{I}_{i,k}\right|=0.
\end{cases}\label{eq:f_approx-1}
\end{align}
Similar to (\ref{eq:algo_sp}), the algorithm is given by 
\begin{align}
a_{i,k+1} & =a_{i,k}+\beta_{k}\widehat{g}_{i,k}^{\left(\textrm{I}\right)}=a_{i,k}+\beta_{k}\Phi_{i,k}\widetilde{f}_{i}^{(\textrm{I})}\left(\boldsymbol{a},\mathbf{S}_{k},\mathcal{I}_{i,k}\right)\label{eq:algo_sp-1}
\end{align}
The basic idea is to consider $\left(N-1\right)\sum_{j\in\mathcal{I}_{i,k}}\widetilde{u}_{j,k}/\left|\mathcal{I}_{i,k}\right|$
as a surrogate function of $\sum_{j\in\mathcal{N}\setminus\left\{ i\right\} }\widetilde{u}_{j,k}$,
in the case where the set $\mathcal{I}_{i,k}$ is non-empty. Otherwise,
node~$i$ does not know any utility of the other nodes, it then cannot
estimate the global utility of the system. As a result, node~$i$
keeps its previous action, \emph{i.e.}, $a_{i,k+1}=a_{i,k}$. Note
that different users may have different knowledge of the global utility
as $\mathcal{I}_{i,k}$ is independent for each node~$i$. For example,
node~$i$ may know $\widetilde{u}_{j,k}$ of a different node~$j$,
whereas node~$j$ may not know $\widetilde{u}_{i,k}$.

\begin{algorithm}[h]
\caption{\label{alg:ESSP-based-Algorithm-1}DOSP algorithm for each node $i$
with incomplete information of the utilities of other nodes}
\begin{enumerate}
\item Initialize $k=0$ and set the action $a_{i,0}$ randomly.
\item Generate a random variable $\Phi_{i,k}$, perform action $a_{i,k}+\gamma_{k}\Phi_{i,k}$.
\item Estimate $\widetilde{u}_{i,k}$, exchange its value with the other
nodes and calculate $\widetilde{f}_{i}^{(\textrm{I})}$ using (\ref{eq:f_approx-1})
based on the collected local utilities.
\item Update $a_{i,k+1}$ according to equation (\ref{eq:algo_sp-1}).
\item $k=k+1$, go to 2.
\end{enumerate}
\end{algorithm}

Due to the additional random term $\mathcal{I}_{i,k}$, the convergence
analysis of Algorithm~\ref{alg:ESSP-based-Algorithm-1} is more complicated
than that of Algorithm~\ref{alg:ESSP-based-Algorithm}. We start
with Lemma~\ref{lem:aver_f_in}, which is useful for the analysis
in what follows
\begin{lem}
\label{lem:aver_f_in}The expected value of \textup{$\widetilde{f}_{i}^{(\textrm{I})}\left(\boldsymbol{a},\mathbf{S}_{k},\mathcal{I}_{i,k}\right)$
}\textup{\emph{over all possible sets}}\textup{ $\mathcal{I}_{i,k}$
}\textup{\emph{is proportional to}}\textup{ $\widetilde{f}\left(\boldsymbol{a},\mathbf{S}_{k}\right)$,
}\textup{\emph{i.e.}}\textup{, 
\begin{equation}
\mathbb{E}_{\mathcal{I}_{i,k}}\left(\widetilde{f}_{i}^{(\textrm{I})}\left(\boldsymbol{a},\mathbf{S}_{k},\mathcal{I}_{i,k}\right)\right)=(1-(1-p)^{N})\widetilde{f}\left(\boldsymbol{a},\mathbf{S}_{k}\right).\label{eq:f_inc}
\end{equation}
}
\end{lem}
\begin{IEEEproof}
See Appendix \ref{subsec:Proof_in}.
\end{IEEEproof}
To simplify the notation, introduce 
\begin{equation}
q=\mathbb{P}\left(\left|\mathcal{I}_{i,k}\right|\neq0\right)=1-(1-p)^{N}.\label{eq:q_def}
\end{equation}
 Similar to (\ref{eq:RM_form}), we rewrite (\ref{eq:algo_sp-1})
as 
\begin{align}
a_{i,k+1} & =a_{i,k}+\alpha_{2}q\beta_{k}\gamma_{k}\left(F_{i}'\left(\boldsymbol{a}_{k}\right)+b_{i,k}^{\left(\textrm{I}\right)}+\frac{e_{i,k}^{\left(\textrm{I}\right)}}{\alpha_{2}q\gamma_{k}}\right)\label{eq:RM_form-1}
\end{align}
with 
\begin{align}
b_{i,k}^{\left(\textrm{I}\right)} & =\frac{\mathbb{E}_{\mathbf{S},\boldsymbol{\Phi},\boldsymbol{\eta},\mathcal{I}_{i,k}}\left(\widehat{g}_{i,k}^{(\textrm{I})}\right)}{\alpha_{2}q\gamma_{k}}-F_{i}'\left(\boldsymbol{a}_{k}\right),\label{eq:bias_in}\\
e_{i,k}^{\left(\textrm{I}\right)} & =\widehat{g}_{i,k}^{\left(\textrm{I}\right)}-\mathbb{E}_{\mathbf{S},\boldsymbol{\Phi},\boldsymbol{\eta},\mathcal{I}_{i,k}}\left(\widehat{\boldsymbol{g}}_{k}^{(\textrm{I})}\right).\label{eq:sto_n_in}
\end{align}

We need to investigate the property of $b_{i,k}^{\left(\textrm{I}\right)}$
and $e_{i,k}^{\left(\textrm{I}\right)}$ in order to show the convergence
of Algorithm~\ref{alg:ESSP-based-Algorithm-1}. The proof is more
complicated because of the additional random term $\mathcal{I}_{i,k}$
in both $b_{i,k}^{\left(\textrm{I}\right)}$ and $e_{i,k}^{\left(\textrm{I}\right)}$
compared with $b_{i,k}$ and $e_{i,k}$ discussed in Section~\ref{sec:Distributed-allocation-algorithm}.

\begin{mythm}\label{prop:converge_rp_in}In the situation where nodes
do not have the access to all the other nodes' local utilities and
the objective function is approximated by applying (\ref{eq:f_approx-1}),
then we still have $\boldsymbol{a}_{k}\rightarrow\boldsymbol{a}^{*}$
as $k\rightarrow\infty$ almost surely by applying Algorithm~\ref{alg:ESSP-based-Algorithm-1},
as long as the assumptions A1-A6 hold and $\left\Vert \boldsymbol{a}_{k}\right\Vert <\infty$
almost surely. \end{mythm}
\begin{IEEEproof}
See Appendix \ref{subsec:Proof-of-convergence_in}.
\end{IEEEproof}
\begin{remark}Although the asymptotic convergence still holds, the
convergence speed is reduced if the information of the objective function
is incomplete. By comparing (\ref{eq:RM_form}) and (\ref{eq:RM_form-1}),
we can see that the equivalent step size is decreased by $q$ times.
Moreover, the variance of the stochastic noise $e_{i,k}^{\left(\textrm{I}\right)}$
is higher, as the randomness is more significant when we use $n<N$
random symbols to represent the average of $N$ symbols. More discussion
related to the convergence rate will be provided in Section~\ref{sec:Convergence-rate}.\end{remark}

\section{Convergence rate \label{sec:Convergence-rate}}

In this section, we study the average convergence rate of the proposed
algorithm in order to investigate how fast the proposed algorithms
converge to the optimum from a quantitative point of view. The analysis
also provides a detailed guideline of setting the parameters $\beta_{k}$
and $\gamma_{k}$ which determine the convergence rate. We start with
the analysis considering general forms of $\beta_{k}$ and $\gamma_{k}$.
A widely used example is then considered afterwards.

\subsection{General analysis}

As a common setting in the analysis of the convergence rate \cite{nemirovski2009robust},
we have an additional assumption on the concavity of the objective
function in this section, \emph{i.e.}, 
\begin{itemize}
\item \emph{A7}. $F\left(\boldsymbol{a}\right)$ is strongly concave, there
exists $\alpha_{5}>0$ such that 
\begin{equation}
\left(\boldsymbol{a}-\boldsymbol{a}^{*}\right)^{T}\nabla F\left(\boldsymbol{a}_{k}\right)\leq-\alpha_{5}\left\Vert \boldsymbol{a}-\boldsymbol{a}^{*}\right\Vert _{2}^{2},\quad\forall\boldsymbol{a}\in\mathcal{A}.\label{eq:order1_2}
\end{equation}
\end{itemize}
In this section, we are interested in the evolution of the average
divergence $D_{k}=\mathbb{E}(\left\Vert \boldsymbol{a}_{k}-\boldsymbol{a}^{*}\right\Vert _{2}^{2}$).
In order to distinguish the two versions of algorithms with complete
and incomplete information of local utilities, we use $D_{k}^{\left(\textrm{C}\right)}$
and $D_{k}^{\left(\textrm{I}\right)}$ to denote the divergence resulted
by Algorithm~\ref{alg:ESSP-based-Algorithm} and Algorithm~\ref{alg:ESSP-based-Algorithm-1},
respectively. 

An essential step of the analysis of the convergence rate is to investigate
the relation of the divergence between two successive iterations. 

Denote $M^{\left(\mathrm{C}\right)}$ and $M^{\left(\mathrm{I}\right)}$
as the upper bounds of $\mathbb{E}(\left\Vert \widehat{\boldsymbol{g}}_{k}\right\Vert _{2}^{2})$
and of $\mathbb{E}(\left\Vert \widehat{\boldsymbol{g}}_{k}^{\left(\mathrm{I}\right)}\right\Vert _{2}^{2})$
in Algorithm~\ref{alg:ESSP-based-Algorithm} and Algorithm~\ref{alg:ESSP-based-Algorithm-1}
respectively. Our result is stated in Lemma~(\ref{lem:inequality_D}). 
\begin{lem}
\label{lem:inequality_D}Assume that A7 holds, for any $k\geq1$,
$D_{k}^{\left(\mathrm{C}\right)}$ resulted by Algorithm~\ref{alg:ESSP-based-Algorithm}
is such that
\begin{align}
 & D_{k+1}^{\left(\mathrm{C}\right)}\leq\left(1-2\alpha_{2}\alpha_{5}\beta_{k}\gamma_{k}\right)D_{k}^{\left(\mathrm{C}\right)}\nonumber \\
 & \qquad\qquad\qquad+N^{\frac{5}{2}}\alpha_{1}\alpha_{3}^{3}\beta_{k}\gamma_{k}^{2}\sqrt{D_{k}^{\left(\mathrm{C}\right)}}+M^{\left(\mathrm{C}\right)}\beta_{k}^{2},\label{eq:dk+1_dk}
\end{align}
Similarly, $D_{k}^{\left(\mathrm{I}\right)}$ resulted by Algorithm~\ref{alg:ESSP-based-Algorithm-1}
is such that 
\begin{align}
 & D_{k+1}^{\left(\mathrm{I}\right)}\leq\left(1-2\alpha_{2}\alpha_{5}q\beta_{k}\gamma_{k}\right)D_{k}^{\left(\mathrm{I}\right)}\nonumber \\
 & \qquad\qquad\qquad+N^{\frac{5}{2}}\alpha_{1}\alpha_{3}^{3}q\beta_{k}\gamma_{k}^{2}\sqrt{D_{k}^{\left(\mathrm{I}\right)}}+M^{\left(\mathrm{I}\right)}\beta_{k}^{2}.\label{eq:dk+1_dk_I}
\end{align}
\end{lem}
\begin{IEEEproof}
See Appendix \ref{subsec:Proof-of-Lemma_D}. 
\end{IEEEproof}
We can see that both (\ref{eq:dk+1_dk}) and (\ref{eq:dk+1_dk_I})
can be written in the simplified general form 
\begin{equation}
D_{k+1}\leq\left(1-A\beta_{k}\gamma_{k}\right)D_{k}+B\beta_{k}\gamma_{k}^{2}\sqrt{D_{k}}+C\beta_{k}^{2},\label{eq:dk+1_dk_0}
\end{equation}
with positive constants
\begin{equation}
A^{\left(\mathrm{C}\right)}=2\alpha_{2}\alpha_{5},\:B^{\left(\mathrm{C}\right)}=N^{\frac{5}{2}}\alpha_{1}\alpha_{3}^{3},\:C^{\left(\mathrm{C}\right)}=M^{\left(\mathrm{C}\right)}\label{eq:ABC}
\end{equation}
in (\ref{eq:dk+1_dk}) and 
\begin{equation}
A^{\left(\mathrm{I}\right)}=2q\alpha_{2}\alpha_{5},\:B^{\left(\mathrm{I}\right)}=qN^{\frac{5}{2}}\alpha_{1}\alpha_{3}^{3},\:C^{\left(\mathrm{I}\right)}=M^{\left(\mathrm{I}\right)}\label{eq:ABCI}
\end{equation}
in (\ref{eq:dk+1_dk_I}). 

\begin{remark}In the constrained optimization problem, we can obtain
the same result as Lemma~\ref{lem:inequality_D} directly from (\ref{eq:divergence_def_1}),
for any $k\geq K_{\mathrm{c}}$. In fact, the unconstrained optimization
problem can be seen as a special case of the constrained optimization
problem with $a_{i,\min}=-\infty$, $a_{i,\max}=+\infty$, and $K_{\mathrm{c}}=0$.
For this reason, we consider the general problem with $K_{\mathrm{c}}$
taken into account, in the rest of this section.\end{remark}

Introduce 
\[
K_{0}=\underset{k\geq K_{\mathrm{c}},\beta_{k}\gamma_{k}<1/A}{\arg\min}k,
\]
which implies that $1-A\beta_{k}\gamma_{k}>0$ and $k\geq K_{\mathrm{c}}$
as $k\geq K_{0}$.

Lemma~\ref{lem:inequality_D} provides us the relation between $D_{k+1}$
and $D_{k}$. Our next goal is to search a vanishing upper bound of
$D_{k}$ using (\ref{eq:dk+1_dk_0}). In other words, we aim to search
a sequence $U_{K_{0}},\ldots,U_{k},\ldots$ such that 
\[
U_{k+1}\leq U_{k}\textrm{ and }D_{k}\leq U_{k},\:\forall k\geq K_{0}.
\]
This type of analysis is usually performed by induction: consider
a given expression of $U_{k}$ and assume that $D_{k}\leq U_{k}$,
one needs to show that $D_{k+1}\leq U_{k+1}$ by applying (\ref{eq:dk+1_dk_0}). 

An important issue is then the proper choice of the form of the upper
bound $U_{k}$. Note that there exists only one step-size $\beta_{k}$
in the classical stochastic approximation algorithm described by (\ref{eq:algo_update}),
it is relatively simple to determine the form of $U_{k}$ with a further
setting $\beta_{k}=\beta_{0}k^{-1}$, see \cite{nemirovski2009robust}.
Our problem is much more complicated as we use two step-sizes $\beta_{k}$
and $\gamma_{k}$ with general form under Assumption \emph{A4}. The
following lemma presents an important property of $U_{k}$.
\begin{lem}
\label{lem:nece_rate}If there exists a decreasing sequence $U_{K_{0}},\ldots,U_{k},\ldots$
such that $D_{k+1}\leq U_{k+1}$ can be deduced from $D_{k}\leq U_{k}$
and (\ref{eq:dk+1_dk_0}), then , 
\begin{equation}
U_{k}\geq\left(\frac{B}{2A}\gamma_{k}+\sqrt{\left(\frac{B}{2A}\right)^{2}\gamma_{k}^{2}+\frac{C}{A}\frac{\beta_{k}}{\gamma_{k}}}\right)^{2}.\label{eq:lower_u}
\end{equation}
\end{lem}
\begin{IEEEproof}
See Appendix~\ref{subsec:Proof-necerate}.
\end{IEEEproof}
Note that the lower bound of $U_{k}$ is vanishing as $\gamma_{k}\rightarrow0$
and $\beta_{k}/\gamma_{k}\rightarrow0$. Such bound means that, using
induction by developing (\ref{eq:dk+1_dk_0}), the convergence rate
of $D_{k}$ cannot be \emph{better} than the decreasing speed of $\gamma_{k}^{2}$
\emph{and} of $\beta_{k}/\gamma_{k}$. After the verification of the
existence of bounded constant $\vartheta$ or $\varrho$ such that
$D_{k}\leq\vartheta^{2}\gamma_{k}^{2}$ or $D_{k}\leq\varrho^{2}\gamma_{k}/\beta_{k}$,
we obtain the final results stated as follows. 

\begin{mythm}\label{prop:rate} Define the following parameters:
\begin{equation}
\chi_{k}=\frac{1-\left(\frac{\gamma_{k+1}}{\gamma_{k}}\right)^{2}}{\beta_{k}\gamma_{k}},\epsilon_{1}=\max_{k\geq K_{0}}\chi_{k},\epsilon_{2}=\max_{k\geq K_{0}}\frac{\beta_{k}}{\gamma_{k}^{3}},\label{eq:chi_w}
\end{equation}
\begin{equation}
\varpi_{k}=\frac{1-\frac{\beta_{k+1}\gamma_{k+1}^{-1}}{\beta_{k}\gamma_{k}^{-1}}}{\beta_{k}\gamma_{k}},\epsilon_{3}=\max_{k\geq K_{0}}\varpi_{k},\epsilon_{4}=\max_{k\geq K_{0}}\sqrt{\frac{\gamma_{k}^{3}}{\beta_{k}}}\label{eq:small_p}
\end{equation}
 If $\chi_{k}<A$ for any $k\geq K_{0}$, then 
\begin{equation}
D_{k}\leq\vartheta^{2}\gamma_{k}^{2},\:\forall k\geq K_{0},\label{eq:Dk_up1}
\end{equation}
with
\begin{equation}
\vartheta\geq\max\left\{ \frac{\sqrt{D_{K_{0}}}}{\gamma_{K_{0}}},\frac{B+\sqrt{B^{2}+4C\epsilon_{2}\left(A-\epsilon_{1}\right)}}{2\left(A-\epsilon_{1}\right)}\right\} ,\label{eq:para_0}
\end{equation}
If $\varpi_{k}<A$ for any $k\geq K_{0}$, then 
\begin{equation}
D_{k}\leq\varrho^{2}\frac{\beta_{k}}{\gamma_{k}},\:\forall k\geq K_{0},\label{eq:Dk_up2}
\end{equation}
with
\begin{equation}
\varrho\geq\max\left\{ \sqrt{\frac{D_{K_{0}}\gamma_{K_{0}}}{\beta_{K_{0}}}},\frac{B\epsilon_{4}+\sqrt{\left(B\epsilon_{4}\right)^{2}+4C\left(A-\epsilon_{3}\right)}}{2\left(A-\epsilon_{3}\right)}\right\} .\label{eq:para_0-1}
\end{equation}
\end{mythm}
\begin{IEEEproof}
See Appendix \ref{subsec:proof_convergencerate}.
\end{IEEEproof}
For general forms of $\beta_{k}$ and $\gamma_{k}$, Theorem~\ref{prop:rate}
provides two upper bounds of the average divergence $D_{k}^{\left(\textrm{C}\right)}$
and $D_{k}^{\left(\textrm{I}\right)}$. The conditions that $\chi_{k}<A$
and $\varpi_{k}<A$ can be checked easily considering any fixed $\beta_{k}$
and $\gamma_{k}$. In the situation where both conditions are satisfied,
we have 
\[
D_{k}\leq\min\left\{ \vartheta^{2}\gamma_{k}^{2},\varrho^{2}\frac{\beta_{k}}{\gamma_{k}}\right\} ,\:\forall k\geq K_{0}.
\]

From Theorem~\ref{prop:rate}, we can see that the decreasing order
of $D_{k}^{\left(\textrm{C}\right)}$ and $D_{k}^{\left(\textrm{I}\right)}$
mainly depend on the step-sizes $\beta_{k}$ and $\gamma_{k}$, the
incompleteness factor $q$ only has the influence on the constant
terms $\vartheta$ and $\varrho$, which are functions of the parameters
$A$, $B$, and $C$ defined in (\ref{eq:ABC}) and (\ref{eq:ABCI}).
The results of convergence rate are useful to properly choose the
parameters of the algorithm. Intuitively, we need to make $\gamma_{k}^{2}$
or $\beta_{k}/\gamma_{k}$ decrease as fast as possible, having the
constant $\epsilon_{1}$, $\epsilon_{2}$, $\epsilon_{3}$, and $\epsilon_{4}$
as small as possible.

\subsection{A special case }

In this section, we consider an example as mentioned in Example~\ref{rem:condition}:
\begin{equation}
\beta_{k}=\beta_{0}\left(k+1\right)^{-\nu_{1}}\textrm{ and }\gamma_{k}=\gamma_{0}\left(k+1\right)^{-\nu_{2}}.\label{eq:betagamma}
\end{equation}
Recall that $0.5<\nu_{1}<1$ and $0<\nu_{2}\leq1-\nu_{1}$ in order
to meet the conditions in \emph{A4}. 

\begin{mythm}\label{thm:rate_example}Consider $\beta_{k}$ and $\gamma_{k}$
with given forms (\ref{eq:betagamma}), if $\beta_{0}\gamma_{0}\geq\max\left\{ 2\nu_{2},\nu_{1}-\nu_{2}\right\} /A$,
then there exists $\Omega<\infty$, such that 
\begin{equation}
D_{k}\leq\Omega\left(k+1\right)^{-\min\left\{ 2\nu_{2},\nu_{1}-\nu_{2}\right\} },\quad\forall k\geq K_{0}.\label{eq:bound_example}
\end{equation}
\end{mythm}
\begin{IEEEproof}
See Appendix~\ref{subsec:Proof-rate_eg}.
\end{IEEEproof}
We can see the explicit impact of each parameter on the upper bound
of the convergence rate from Theorem~\ref{thm:rate_example}. It
is easy to show that 
\[
\max\left\{ 2\nu_{2},\nu_{1}-\nu_{2}\right\} \leq0.5
\]
with the equality holds when $\nu_{1}=0.75$ and $\nu_{2}=0.25$,
which corresponds to the best choice of $\nu_{1}$ and $\nu_{2}$
to optimize the decreasing order of the upper bound of $D_{k}$. Theorem~\ref{thm:rate_example}
also provides a \emph{sufficient} \emph{condition} on the parameters
that should be satisfied in order to ensure the validation of the
convergence rate. 

In what follows, we present a toy example to verify Theorem~~\ref{thm:rate_example}.

\begin{figure}[h]
\begin{centering}
\includegraphics[width=0.98\columnwidth]{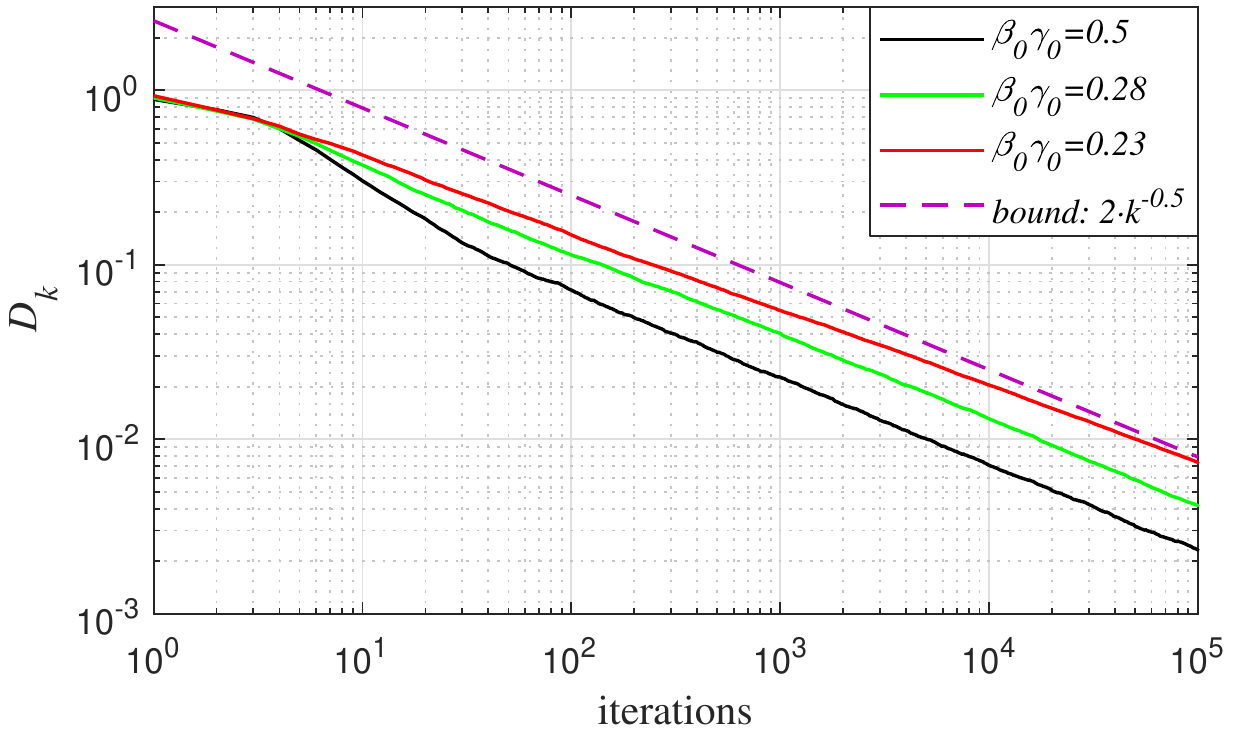} 
\par\end{centering}
\caption{Comparison of the theoretical upper bound $2\left(k+1\right)^{-0.5}$
with the evolution of the average divergence $D_{k}$ (by 2000 simulations)
using $\beta_{k}=\beta_{0}\left(k+1\right)^{0.75}$ and $\gamma_{k}=\left(k+1\right)^{0.25}$,
with $\beta_{0}\in\left\{ 0.23,0.28,0.5\right\} $. \label{fig:verification1} }
\end{figure}

\begin{figure}[h]
\begin{centering}
\includegraphics[width=0.98\columnwidth]{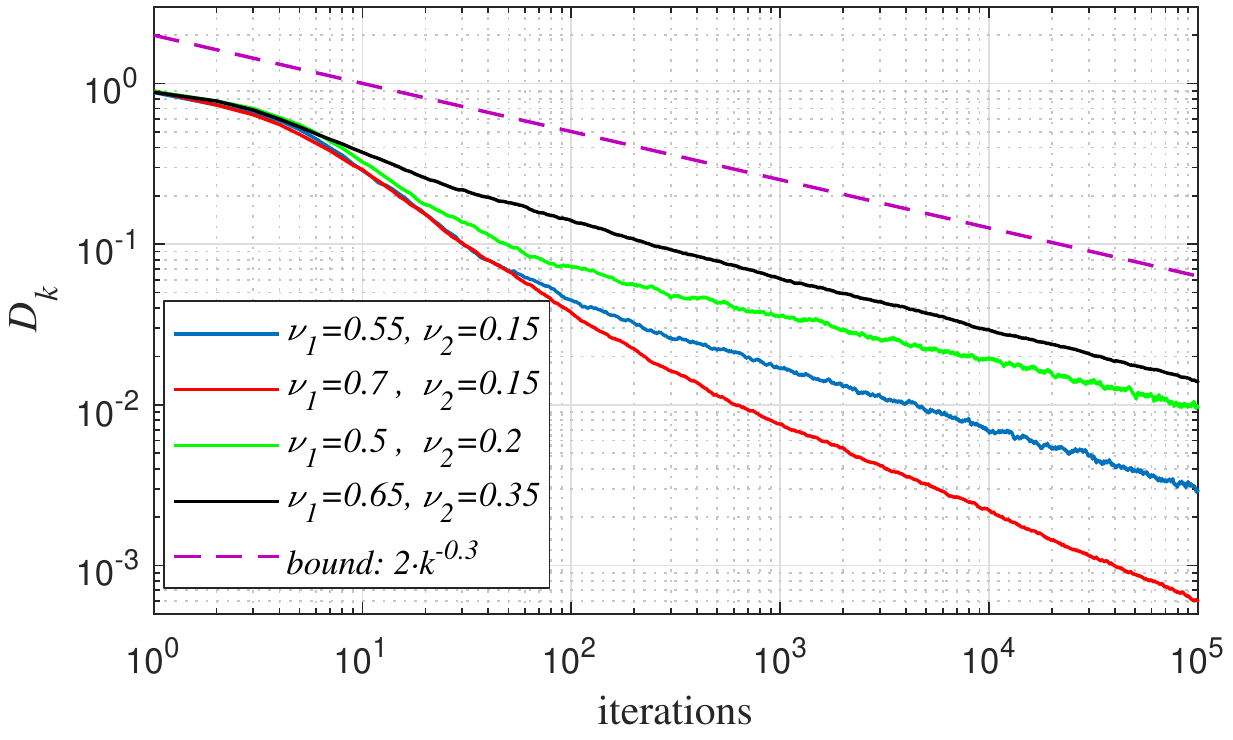} 
\par\end{centering}
\caption{Comparison of the theoretical upper bound $2\left(k+1\right)^{-0.3}$
with the evolution of the average divergence $D_{k}$ (by 2000 simulations)
using $\beta_{k}=0.4\cdot\left(k+1\right)^{\nu_{1}}$ and $\gamma_{k}=\left(k+1\right)^{\nu_{2}}$,
with $\left(\nu_{1},\nu_{2}\right)\in\left\{ \left(0.55,0.15\right),\left(0.7,0.15\right),\left(0.5,0.2\right),\left(0.65,0.35\right)\right\} $\label{fig:verification2} }
\end{figure}

\begin{eg} Consider $N=2$ and a simple quadratic function $f\left(\boldsymbol{a},\mathbf{S}\right)=-s_{1}a_{1}^{2}-s_{2}a_{2}^{2}+a_{1}a_{2}+a_{1}+a_{2}$
with $a_{1},a_{2}\in\left[0,3\right]$. Both $s_{1}$ and $s_{2}$
are realizations of a uniform distributed random variable $S$ which
takes value from an interval $[0.5,1.5]$. Taking the average, the
objective function is 
\[
F\left(\boldsymbol{a}\right)=-a_{1}^{2}-a_{2}^{2}+a_{1}a_{2}+a_{1}+a_{2}.
\]
It is straightforward to get that $F\left(\boldsymbol{a}\right)$
takes its maximum value when $a_{1}=a_{2}=1$. We can also deduce
that $F$ is strongly concave with $\alpha_{5}=1$ in (\ref{eq:order1_2}).
We set the perturbation $\boldsymbol{\Phi}_{k}$ as introduced in
Example~\ref{rem:condition}, with $\alpha_{2}=1$. Thus we have
$A=2\alpha_{2}\alpha_{5}=2$ by (\ref{eq:ABC}). Let $\beta_{k}=\beta_{0}\left(k+1\right)^{-0.75}$
and $\gamma_{k}=\left(k+1\right)^{-0.25}$.

By setting $\beta_{0}\in\left\{ 0.23,0.28,0.5\right\} $, we can verify
the importance of the condition $\beta_{0}\gamma_{0}\geq0.25=\max\left\{ 2\nu_{2},\nu_{1}-\nu_{2}\right\} /A$
as presented in Theorem~\ref{thm:rate_example}. Figure~\ref{fig:verification1}
shows the comparison results between the divergence $D_{k}$ averaged
by 1000 simulations and the bound $\Omega\left(k+1\right)^{-0.5}$
from (\ref{eq:bound_example}). Note that we are mainly interested
in the asymptotic decreasing order of $D_{k}$, thus we set the constant
term $\Omega=2$ mainly to facilitate the visual comparison of different
curves. We find that the curves of $D_{k}$ decrease not-less-slowly
than the bound as $\beta_{0}\in\left\{ 0.28,0.5\right\} $. When $\beta_{0}=0.23<0.25$,
the convergence rate cannot be guaranteed, which further justify our
claim in Theorem~\ref{thm:rate_example}. 

Now we fix the values of $\beta_{0}$ and $\gamma_{0}$ and consider
$\left(\nu_{1},\nu_{2}\right)\in\{\left(0.55,0.15\right),\left(0.7,0.15\right),\left(0.5,0.2\right),\left(0.65,0.35\right)\}$.
Notice that all the four pairs of $\left(\nu_{1},\nu_{2}\right)$
lead to the same decreasing order of the upper bound (\ref{eq:bound_example}),
as $\min\left\{ 2\nu_{2},\nu_{1}-\nu_{2}\right\} =0.3$. The curves
of $D_{k}$ (averaged by 2000 simulations) are compared with the upper
bound $\Omega\left(k+1\right)^{-0.3}$ in Figure~\ref{fig:verification2}.
Clearly, when the number of iterations is large enough, all the curves
decrease with the same speed or even faster, compared with the bound.\end{eg}


\section{Simulation Results\label{sec:Simulation-Results}}

In this section, we apply our algorithm to a power control problem
as introduced in Section~\ref{sec:Motivating-Example} in order to
have some numerical results. We consider (\ref{eq:u_function}) as
the local utility function of each node. The time varying channel
$h_{ij}$ between node~$i$ (transmitter) and node~$j$ (receiver)
is generated using a Gaussian distribution with variance $\sigma_{ii}^{2}=1$
and $\sigma_{ij}^{2}=0.1$, $\forall i\neq j$. Notice that the channel
gain is $s_{ij}=\left|h_{ij}\right|^{2}$. Besides, we set $\sigma^{2}=0.2$,
$\omega=20$ and $\kappa=1$. 

\subsection{Simulations with 4 nodes}

In this section, we consider $N=4$ nodes. In all the simulations
(applying different algorithms), the step size follows $\beta_{k}=2.5k^{-0.75}$,
and the initial values of $a_{0,i}$ ($\forall i\in\mathcal{N}$)
are generated uniformly in the interval $\left(0,20\right]$. In both
Algorithm~\ref{alg:ESSP-based-Algorithm} and Algorithm~\ref{alg:ESSP-based-Algorithm-1},
$\gamma_{k}=12k^{-0.25}$ and $\Phi_{i,k}$ follows the symmetrical
Bernoulli distribution, \emph{i.e.}, $\mathbb{P}\left(\Phi_{i,k}=1\right)=\mathbb{P}\left(\Phi_{i,k}=-1\right)=0.5$,
$\forall k,i$.

We first compare our proposed algorithm with the sine perturbation
based algorithm in \cite{hanif2012convergence}, considering the situation
in which every node has access to all the other nodes' local utilities.
The sine perturbation based algorithm has a similar shape as our stochastic
perturbation algorithm, with the perturbation term $\Phi_{i,k}$ replaced
by a sine function $\lambda_{i}\sin\left(\Omega_{i}t_{k}+\phi_{i}\right)$
where $t_{k}=\sum_{k'=1}^{k}\beta_{k'}$, $\Omega_{i}\neq\Omega_{i'}$
and $\Omega_{i'}+\Omega_{i}\neq\Omega_{i''}$ $\forall i,i',i''$.
In the simulation, we set $\Omega_{1}=63$, $\Omega_{2}=70$, $\Omega_{3}=56$,
$\Omega_{4}=49$, $\lambda_{i}=1.5$ and $\phi_{i}=0$, $\forall i\in\mathcal{N}$.
Notice that this algorithm is not easy to implement in practice as
it is hard to choose all the parameters properly, especially when
$\mathcal{\left|N\right|}$ is large.

Furthermore, in order to show the efficiency of our algorithm, we
simulate also an ideal algorithm using the exact gradient calculated
by (\ref{eq:deri_theo}) which is costly to be obtained in practice
as discussed in Section~\ref{sec:Motivating-Example}. 

\begin{figure}[h]
\begin{centering}
\includegraphics[width=0.98\columnwidth]{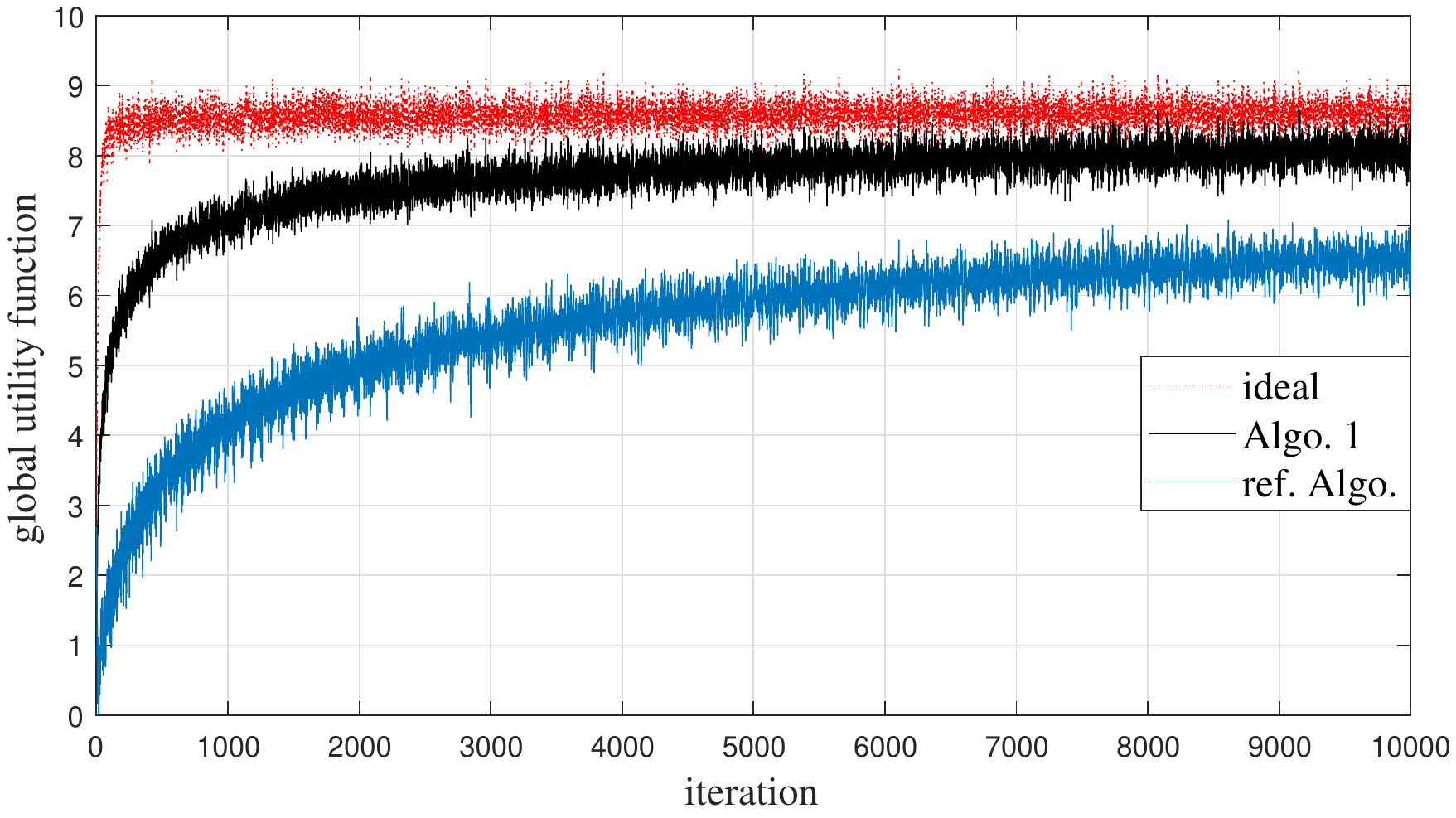} 
\par\end{centering}
\caption{Evolution of the utility function, average results by 500 simulations\label{fig:aver_u} }
\end{figure}

\begin{figure}[h]
\begin{centering}
\includegraphics[width=0.98\columnwidth]{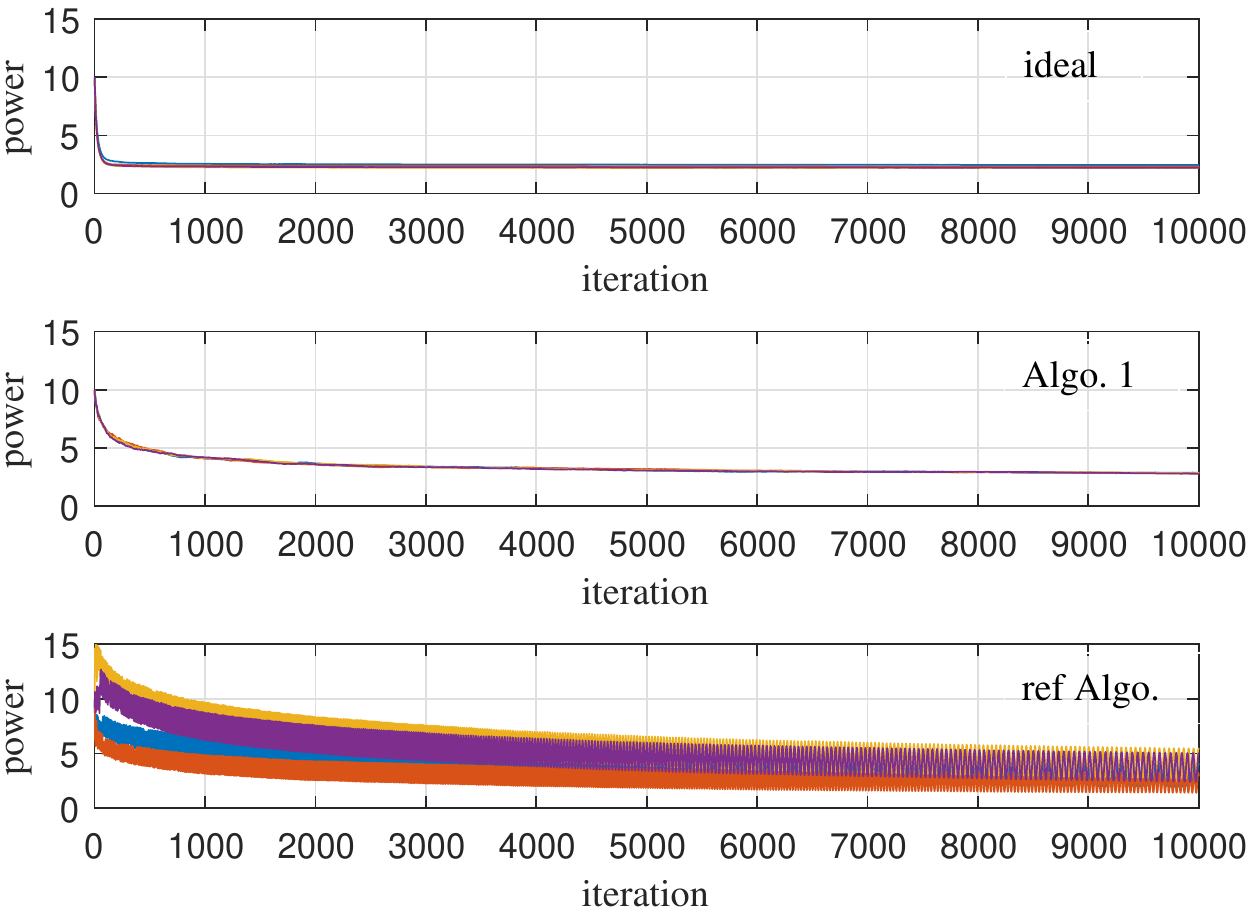} 
\par\end{centering}
\caption{Evolution of power (action) of 4 nodes, average results by 500 simulations\label{fig:aver_a} }
\end{figure}

We have performed 500 independent simulations to obtain the average
results shown in Figures \ref{fig:aver_u} and \ref{fig:aver_a}.
Figure \ref{fig:aver_u} represents the utility function $f\left(\boldsymbol{a},\mathbf{S}\right)/N$
as a function of number of iterations. We find that our algorithm
converges faster than the reference algorithm proposed in \cite{hanif2012convergence}
\footnote{The reference algorithm is quite sensitive to the parameters, the
presented results are the best that we have found so far.}. Figure \ref{fig:aver_a} shows the evolution of the power (action)
of the four nodes. Notice that the four curves representing the action
of each node are close in average in each sub-figure, since we consider
the model with symmetric parameters. We find the oscillation of the
power (action) is more significant by applying the reference algorithm.

\begin{figure}[h]
\begin{centering}
\includegraphics[width=0.98\columnwidth]{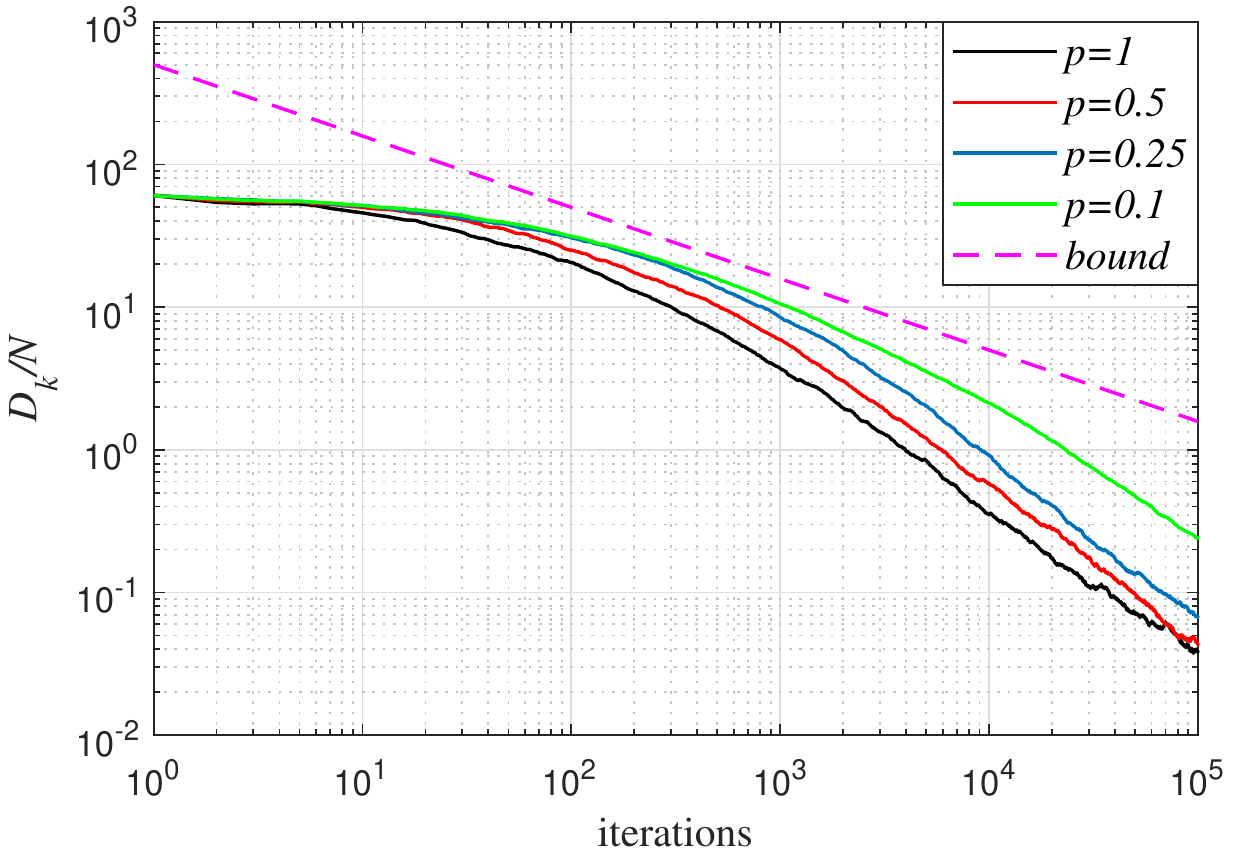} 
\par\end{centering}
\caption{Evolution of $D_{k}/N$, average results by 100 simulations, with
$p\in\left\{ 1,0.5,0.25,0.1\right\} $.\label{incomp}}
\end{figure}

Now we consider the situation where each node has incomplete collection
of local utilities. We perform 100 independent simulation with $p\in\left\{ 1,0.5,0.25,0.1\right\} $.
Recall that $p$ defined in (\ref{eq:p}) represents the level of
incompleteness. The results are shown in Figure~\ref{incomp}. We
can see that the convergence speed decreases as the value of $p$
goes smaller. Such influence is not significant even if $p=0.5$,
$i.e.$, a node has only $50\%$ chance to know the local utility
of another user, which reduces a lot the information exchange in the
network. As the value of $p$ is very small, \emph{i.e.}, $p=0.1$,
same trend of convergence can be observed, although the algorithm
converges slowly. On the figure, we show the results obtained after
up to $10^{4}$ iterations only, which explains why for $p=0.1$ the
algorithm has not converged yet to the optimal solution.

\subsection{Simulations with 10 nodes}

\begin{figure}[h]
\begin{centering}
\includegraphics[width=0.98\columnwidth]{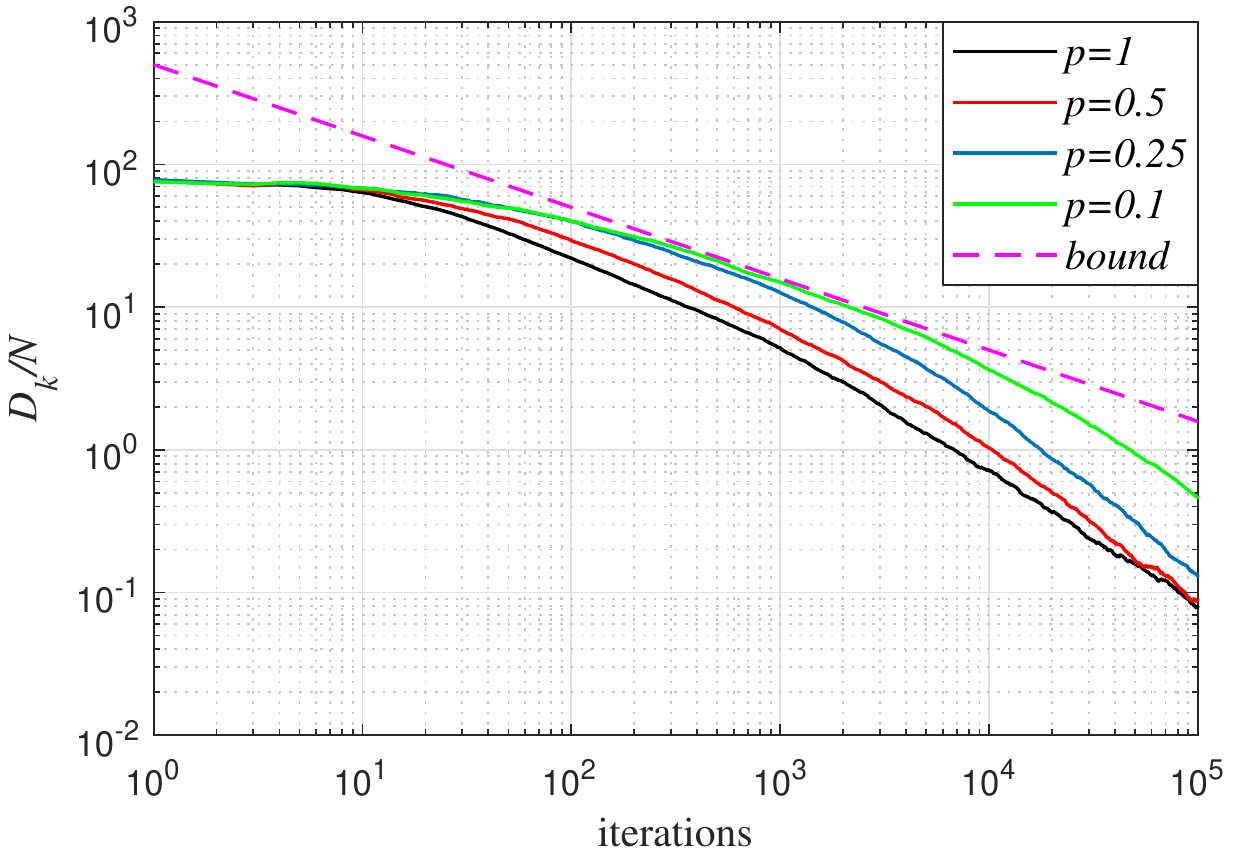} 
\par\end{centering}
\caption{Evolution of $D_{k}/N$, average results by 100 simulations,with $p\in\left\{ 1,0.5,0.25,0.1\right\} $.\label{incomp-1}}
\end{figure}

In this section, we consider a more challenging case with $N=10$.
We choose $\beta_{k}=2\left(k+1\right)^{-0.75}$ and $\gamma_{k}=12\left(k+1\right)^{-0.25}$.
The results are presented in Figure \ref{incomp-1} with $p\in\left\{ 1,0.5,0.25,0.1\right\} $,
note that we have also plotted a line representing the optimum value
of the average utility function. The shape of the curves in Figure~\ref{incomp-1}
is similar to those in Figure~\ref{incomp}. The algorithm converges
slower as the number of nodes increases, yet the influence of the
incompleteness of the received local utilities is less important.

\section{Conclusion\label{sec:Conclusion}}

In this paper we have a challenging distributed optimization problems,
under the assumption that only a numerical value of the stochastic
state-dependent local utility function of the node is available at
each time and nodes need to exchange their local values to optimize
the total utilities of the network. We have developed fully distributed
algorithms that converge to the optimum, in the situations where each
node has the knowledge of all or only a part of the local utilities
of the others. The convergence of our algorithms is examined by studying
our algorithm using stochastic approximation technique. The convergence
rate of the algorithms are derived. Numerical results are also provided
for illustration.

\appendix
%dummy comment inserted by tex2lyx to ensure that this paragraph is not empty%dummy comment inserted by tex2lyx to ensure that this paragraph is not empty%dummy comment inserted by tex2lyx to ensure that this paragraph is not empty

\subsection{Proof of Lemma~\ref{lem:bias}}

\label{subsec:Proof-bias}In this proof, we mainly need to find the
relation between $\overline{\boldsymbol{g}}_{k}$ (the expected estimation
of gradient) and $\nabla F\left(\boldsymbol{a}_{k}\right)$ (the actual
gradient of the objective function), so that the upper bound of $\left\Vert \boldsymbol{b}_{k}\right\Vert $
can be derived.

From the definition of $\overline{\boldsymbol{g}}_{k}$, we have
\begin{align}
\overline{\boldsymbol{g}}_{k} & =\mathbb{E}_{\mathbf{S},\boldsymbol{\Phi},\boldsymbol{\eta}}\left(\boldsymbol{\Phi}_{k}\left(f\left(\boldsymbol{a}_{k}+\gamma_{k}\boldsymbol{\Phi}_{k},\mathbf{S}_{k}\right)+\sum_{i\in\mathcal{N}}\eta_{i,k}\right)\right)\nonumber \\
 & =\mathbb{E}_{\boldsymbol{\Phi}}\left(\boldsymbol{\Phi}_{k}\mathbb{E}_{\mathbf{S}}\left(f\left(\boldsymbol{a}_{k}+\gamma_{k}\boldsymbol{\Phi}_{k},\mathbf{S}_{k}\right)\right)\right)\nonumber \\
 & =\mathbb{E}_{\boldsymbol{\Phi}}\left(\boldsymbol{\Phi}_{k}F\left(\boldsymbol{a}_{k}+\gamma_{k}\boldsymbol{\Phi}_{k}\right)\right),\label{eq:rp_1}
\end{align}
recall that the additive noise $\eta_{i,k}$ is zero mean and $F$
is the expected value of $f$ by definition.

Based on Taylor's theorem and mean-valued theorem, there exists $\widetilde{\boldsymbol{a}}_{k}$
locating between $\boldsymbol{a}_{k}$ and $\boldsymbol{a}_{k}+c_{k}\boldsymbol{\Phi}_{k}$
such that 
\begin{align}
 & F\left(\boldsymbol{a}_{k}+\gamma_{k}\boldsymbol{\Phi}_{k}\right)=F\left(\boldsymbol{a}_{k}\right)+\!\sum_{j\in\mathcal{N}}\!\gamma_{k}\Phi_{j,t}F_{j}'\left(\boldsymbol{a}_{k}\right)\nonumber \\
 & \qquad\qquad\qquad+\sum_{j_{1},j_{2}\in\mathcal{N}}\frac{\gamma_{k}^{2}}{2}\Phi_{j_{1},k}\Phi_{j_{2},k}F_{j_{1},j_{2}}''\left(\widetilde{\boldsymbol{a}}_{k}\right).\label{eq:rp_2}
\end{align}
Benefit from the properties of $\boldsymbol{\Phi}_{k}$ (\emph{A4}),
we have, $\forall i\in\mathcal{N}$,
\begin{align}
\overline{g}_{i,k} & =F\left(\boldsymbol{a}_{k}\right)\mathbb{E}_{\boldsymbol{\Phi}}\left(\Phi_{i,k}\right)+\gamma_{k}\sum_{j\in\mathcal{N}}F_{j}'\left(\boldsymbol{a}_{k}\right)\mathbb{E}_{\boldsymbol{\Phi}}\left(\Phi_{i,k}\Phi_{j,k}\right)\nonumber \\
 & \quad\qquad+\sum_{j_{1},j_{2}\in\mathcal{N}}\frac{\gamma_{k}^{2}}{2}\mathbb{E}_{\boldsymbol{\Phi}}\left(\Phi_{i,k}\Phi_{j_{1},k}\Phi_{j_{2},k}F_{j_{1},j_{2}}''\left(\widetilde{\boldsymbol{a}}_{k}\right)\right)\nonumber \\
 & =\alpha_{2}\gamma_{k}\left(F_{i}'\left(\boldsymbol{a}_{k}\right)+b_{i,k}\right),\label{eq:g_aver1}
\end{align}
from (\ref{eq:rp_1}) and (\ref{eq:rp_2}), with 
\[
b_{i,k}=\sum_{j_{1},j_{2}\in\mathcal{N}}\frac{\gamma_{k}}{2\alpha_{2}}\mathbb{E}_{\boldsymbol{\Phi}}\left(\Phi_{i,k}\Phi_{j_{1},k}\Phi_{j_{2},k}F_{j_{1},j_{2}}''\left(\widetilde{\boldsymbol{a}}_{k}\right)\right).
\]
From assumptions \emph{A2} and \emph{A4}, $\left|b_{i,k}\right|$
can be upper bounded by
\begin{align}
\left|b_{i,k}\right| & \leq\sum_{j_{1},j_{2}\in\mathcal{N}}\frac{\gamma_{k}}{2\alpha_{2}}\mathbb{E}_{\boldsymbol{\Phi}}\left(\left|\Phi_{i,k}\right|\cdot\left|\Phi_{j_{1},k}\right|\cdot\left|\Phi_{j_{2},k}\right|\cdot\left|F_{j_{1},j_{2}}''\left(\widetilde{\boldsymbol{a}}_{k}\right)\right|\right)\nonumber \\
 & \leq\sum_{j_{1},j_{2}\in\mathcal{N}}\frac{\gamma_{k}}{2\alpha_{2}}\mathbb{E}_{\boldsymbol{\Phi}}\left(\alpha_{3}^{3}\alpha_{1}\right)\nonumber \\
 & =\gamma_{k}N^{2}\frac{\alpha_{3}^{3}\alpha_{1}}{2\alpha_{2}},\label{eq:bound_d}
\end{align}
then we can get (\ref{eq:bias_bound}). Thus $\left\Vert \boldsymbol{b}_{k}\right\Vert \rightarrow0$
as $\gamma_{k}$ is vanishing, which concludes the proof.

\subsection{Proof of Lemma~\ref{lem:stochastic_noise}}

\label{subsec:Proof-e}The proof of Lemma~\ref{lem:stochastic_noise}
is mainly by the application of Doob's martingale inequality \cite{doob1953stochastic}.

We first need to show that the sequence $\left\{ \sum_{k=K}^{K'}\beta_{k}\boldsymbol{e}_{k}\right\} _{K'\geq K}$
is martingale, which is straightforward as $\widehat{\boldsymbol{g}}_{k}$
and $\widehat{\boldsymbol{g}}_{k'}$ are independent if $k\neq k'$
and
\[
\mathbb{E}_{\mathbf{S},\boldsymbol{\Phi},\boldsymbol{\eta}}\left(\boldsymbol{e}_{k}\right)=\mathbb{E}_{\mathbf{S},\boldsymbol{\Phi},\boldsymbol{\eta}}\left(\widehat{\boldsymbol{g}}_{k}-\mathbb{E}_{\mathbf{S},\boldsymbol{\Phi},\boldsymbol{\eta}}\left(\widehat{\boldsymbol{g}}_{k}\right)\right)=\mathbf{0}.
\]
Then, by Doob's martingale inequality, for any positive constant $\rho$,
we have
\begin{align}
 & \mathbb{P}\left(\sup_{K'\geq K}\left\Vert \sum_{k=K}^{K'}\beta_{k}\boldsymbol{e}_{k}\right\Vert \geq\rho\right)\nonumber \\
 & \leq\frac{1}{\rho^{2}}\mathbb{E}_{\mathbf{S},\boldsymbol{\Phi},\boldsymbol{\eta}}\left(\left\Vert \sum_{k=K}^{K'}\beta_{k}\boldsymbol{e}_{k}\right\Vert ^{2}\right)\nonumber \\
 & =\frac{1}{\rho^{2}}\mathbb{E}_{\mathbf{S},\boldsymbol{\Phi},\boldsymbol{\eta}}\left(\sum_{k=K}^{K'}\sum_{k'=K}^{K'}\beta_{k}\beta_{k'}\boldsymbol{e}_{k}^{T}\cdot\boldsymbol{e}_{k'}\right)\nonumber \\
 & \overset{(a)}{=}\frac{1}{\rho^{2}}\mathbb{E}_{\mathbf{S},\boldsymbol{\Phi},\boldsymbol{\eta}}\left(\sum_{k=K}^{K'}\left\Vert \beta_{k}\boldsymbol{e}_{k}\right\Vert ^{2}\right)\nonumber \\
 & \leq\frac{1}{\rho^{2}}\sum_{k=K}^{\infty}\mathbb{E}_{\mathbf{S},\boldsymbol{\Phi},\boldsymbol{\eta}}\left(\beta_{k}^{2}\left\Vert \widehat{\boldsymbol{g}}_{k}-\mathbb{E}_{\mathbf{S},\boldsymbol{\Phi},\boldsymbol{\eta}}\left(\widehat{\boldsymbol{g}}_{k}\right)\right\Vert ^{2}\right)\nonumber \\
 & =\frac{1}{\rho^{2}}\sum_{k=K}^{\infty}\beta_{k}^{2}\left(\mathbb{E}_{\mathbf{S},\boldsymbol{\Phi},\boldsymbol{\eta}}\left(\left\Vert \widehat{\boldsymbol{g}}_{k}\right\Vert ^{2}\right)-\left\Vert \mathbb{E}_{\mathbf{S},\boldsymbol{\Phi},\boldsymbol{\eta}}\left(\widehat{\boldsymbol{g}}_{k}\right)\right\Vert ^{2}\right)\nonumber \\
 & \leq\frac{1}{\rho^{2}}\sum_{k=K}^{\infty}\beta_{k}^{2}\mathbb{E}_{\mathbf{S},\boldsymbol{\Phi},\boldsymbol{\eta}}\left(\left\Vert \widehat{\boldsymbol{g}}_{k}\right\Vert ^{2}\right)\nonumber \\
 & \overset{\left(b\right)}{\leq}\frac{M}{\rho^{2}}\sum_{k=K}^{\infty}\beta_{k}^{2}\label{eq:e_up_1}
\end{align}
where $(a)$ holds as $\mathbb{E}\left(\boldsymbol{e}_{k}^{T}\cdot\boldsymbol{e}_{k'}\right)=0$
for any $k\neq k'$ and $(b)$ is by Lemma~\ref{lem:bound_gs} as
stated and proved in the end of this section. Since $\lim_{K\rightarrow\infty}\sum_{k=K}^{\infty}\beta_{k}^{2}=0$
by Assumption~\emph{A4 }and $M$ is bounded, we have $\frac{M}{\rho^{2}}\sum_{k=K}^{\infty}\beta_{k}^{2}$
vanishing for any bounded constant $\rho$. Therefore, the probability
that $\left\Vert \sum_{k=K}^{K'}\beta_{k}\boldsymbol{e}_{k}\right\Vert \geq\rho$
is also vanishing, which concludes the proof.

\begin{lem}
\label{lem:bound_gs}If all the conditions in A1-A5 are satisfied
and $\left\Vert \boldsymbol{a}_{k}\right\Vert <\infty$ almost surely,
then there exists a bounded constant $M>0$, such that $\mathbb{E}_{\mathbf{S},\boldsymbol{\Phi},\boldsymbol{\eta}}\left(\left\Vert \widehat{\boldsymbol{g}}_{k}\right\Vert ^{2}\right)<M$
almost surely.
\end{lem}
\begin{IEEEproof}
For any $i\in\mathcal{N}$, we evaluate 
\begin{align}
 & \mathbb{E}_{\mathbf{S},\boldsymbol{\Phi},\boldsymbol{\eta}}\left(\widehat{g}_{i,k}^{2}\right)\nonumber \\
 & =\mathbb{E}_{\mathbf{S},\boldsymbol{\Phi},\boldsymbol{\eta}}\left(\left(\Phi_{i,k}f\left(\boldsymbol{a}_{k}+\gamma_{k}\boldsymbol{\Phi}_{k},\mathbf{S}_{k}\right)+\Phi_{i,k}\sum_{j\in\mathcal{N}}\eta_{j,k}\right)^{2}\right)\nonumber \\
 & \overset{(a)}{=}\mathbb{E}_{\mathbf{S},\boldsymbol{\Phi}}\left(\left(\Phi_{i,k}f\left(\boldsymbol{a}_{k}+\gamma_{k}\boldsymbol{\Phi}_{k},\mathbf{S}_{k}\right)\right)^{2}\right)+N\alpha_{2}\alpha_{4}\nonumber \\
 & \overset{(b)}{\leq}\alpha_{3}^{2}\mathbb{E}_{\mathbf{S},\boldsymbol{\Phi}}\left(\left(f\left(\boldsymbol{a}_{k}+\gamma_{k}\boldsymbol{\Phi}_{k},\mathbf{S}_{k}\right)\right)^{2}\right)+N\alpha_{2}\alpha_{4}\nonumber \\
 & \overset{(c)}{\leq}\alpha_{3}^{2}\mathbb{E}_{\mathbf{S},\boldsymbol{\Phi}}\!\left(\!\left(\left\Vert f\left(\mathbf{0},\mathbf{S}_{k}\right)\right\Vert +\!NL_{\mathbf{S}_{k}}\!\left\Vert \boldsymbol{a}_{k}+\gamma_{k}\boldsymbol{\Phi}_{k}\right\Vert \right)^{2}\!\right)\!+N\alpha_{2}\alpha_{4}\nonumber \\
 & \overset{(d)}{\leq}2\alpha_{3}^{2}\mathbb{E}_{\mathbf{S}}\left(\mu_{\mathbf{S}_{k}}^{2}+N^{2}L_{\mathbf{S}_{k}}^{2}\left(\left\Vert \boldsymbol{a}_{k}\right\Vert +\gamma_{k}N^{\frac{1}{2}}\alpha_{3}\right)^{2}\right)+N\alpha_{2}\alpha_{4}\nonumber \\
 & \overset{(e)}{=}2\alpha_{3}^{2}\left(\mu+N^{2}L\left(\left\Vert \boldsymbol{a}_{k}\right\Vert +\gamma_{k}N^{\frac{1}{2}}\alpha_{3}\right)^{2}\right)+N\alpha_{2}\alpha_{4}\nonumber \\
 & <\infty\label{eq:up_2}
\end{align}
where $(a)$ is due to $\mathbb{E}_{\boldsymbol{\Phi},\boldsymbol{\eta}}\left(\left(\Phi_{i,k}\sum_{j\in\mathcal{N}}\eta_{j,k}\right)^{2}\right)=N\alpha_{2}\alpha_{4}$
and $(b)$ is by Assumption \emph{A4}. From (\ref{eq:Lipschitz_f}),
we have 
\[
\left\Vert f\left(\boldsymbol{a},\mathbf{S}\right)\right\Vert -\left\Vert f\left(\boldsymbol{0},\mathbf{S}\right)\right\Vert \leq\left\Vert f\left(\boldsymbol{a},\mathbf{S}\right)-f\left(\boldsymbol{0},\mathbf{S}\right)\right\Vert \leq NL_{\mathbf{S}}\left\Vert \boldsymbol{a}\right\Vert ,
\]
so $\left\Vert f\left(\boldsymbol{a},\mathbf{S}\right)\right\Vert \leq\left\Vert f\left(\boldsymbol{0},\mathbf{S}\right)\right\Vert +NL_{\mathbf{S}}\left\Vert \boldsymbol{a}\right\Vert $,
$(c)$ can be obtained. We denote $\mu_{\mathbf{S}_{k}}=\left\Vert f\left(\mathbf{0},\mathbf{S}_{k}\right)\right\Vert $
in $(d)$ and the inequality is because of $(x+y)/2\leq\sqrt{\left(x^{2}+y^{2}\right)/2}$,
$\forall x,y\in\mathbb{R}$. In $(e)$, we introduce $\mu=\mathbb{E}_{\mathbf{S}}\left(\mu_{\mathbf{S}_{k}}^{2}\right)$
and $L=\mathbb{E}_{\mathbf{S}}\left(L_{\mathbf{S}_{k}}^{2}\right)$. 

Based on (\ref{eq:up_2}), we see that $\mathbb{E}\left(\left\Vert \widehat{\boldsymbol{g}}_{k}\right\Vert ^{2}\right)=\sum_{i\in\mathcal{N}}\mathbb{E}\left(\widehat{g}_{i,k}^{2}\right)$
is also bounded, which concludes the proof.
\end{IEEEproof}

\subsection{Proof of Theorem \ref{prop:converge_rp}}

\label{subsec:proof_conv}In this proof, we start with the evolution
of the divergence $d_{k}$ as defined in (\ref{eq:divergence_def}),
then we show that $d_{k}$ should be vanishing almost surely by applying
Lemma~\ref{lem:bias} and Lemma~\ref{lem:stochastic_noise}.

By definition, we have
\begin{align}
d_{k+1} & =\left\Vert \boldsymbol{a}_{k+1}-\boldsymbol{a}^{*}\right\Vert ^{2}=\left\Vert \boldsymbol{a}_{k}+\beta_{k}\widehat{\boldsymbol{g}}_{k}-\boldsymbol{a}^{*}\right\Vert ^{2}\nonumber \\
 & =d_{k}+\beta_{k}^{2}\left\Vert \widehat{\boldsymbol{g}}_{k}\right\Vert ^{2}+2\beta_{k}\left(\boldsymbol{a}_{k}-\boldsymbol{a}^{*}\right)^{T}\cdot\widehat{\boldsymbol{g}}_{k}.\label{eq:divergence_n+1_n}
\end{align}
We sum both sides of (\ref{eq:divergence_n+1_n}) to obtain 
\begin{align}
d_{K+1} & =d_{0}+\sum_{k=0}^{K}\left(\beta_{k}^{2}\left\Vert \widehat{\boldsymbol{g}}_{k}\right\Vert ^{2}+2\beta_{k}\left(\boldsymbol{a}_{k}-\boldsymbol{a}^{*}\right)^{T}\cdot\widehat{\boldsymbol{g}}_{k}\right).\label{eq:c0}
\end{align}
Consider (\ref{eq:RM_form}), we can rewrite (\ref{eq:c0}) as 
\begin{align}
d_{K+1} & =d_{0}+\sum_{k=0}^{K}\beta_{k}^{2}\left\Vert \widehat{\boldsymbol{g}}_{k}\right\Vert ^{2}+2\sum_{k=0}^{K}\beta_{k}\left(\boldsymbol{a}_{k}-\boldsymbol{a}^{*}\right)^{T}\cdot\boldsymbol{e}_{k}\nonumber \\
 & +2\alpha_{2}\sum_{k=0}^{K}\beta_{k}\gamma_{k}\left(\boldsymbol{a}_{k}-\boldsymbol{a}^{*}\right)^{T}\cdot\left(\nabla F\left(\boldsymbol{a}_{k}\right)+\boldsymbol{b}_{k}\right).\label{eq:c1}
\end{align}

As stated in Lemma \ref{lem:stochastic_noise}, we have $\lim_{K\rightarrow\infty}\left\Vert \sum_{k=0}^{K}\beta_{k}\boldsymbol{e}_{k}\right\Vert <\infty$
(a.s.), besides $\left\Vert \boldsymbol{a}_{k}-\boldsymbol{a}^{*}\right\Vert <\infty$
a.s., thus 
\begin{equation}
\lim_{K\rightarrow\infty}\left\Vert \sum_{k=0}^{K}\beta_{k}\left(\boldsymbol{a}_{k}-\boldsymbol{a}^{*}\right)^{T}\cdot\boldsymbol{e}_{k}\right\Vert <\infty,\quad\textrm{a.s.}\label{eq:c2}
\end{equation}
From Lemma~\ref{lem:bound_gs}, we have $\left\Vert \widehat{\boldsymbol{g}}_{k}\right\Vert ^{2}<\infty$
almost surely. Then, by \emph{A5}, the following holds almost surely
as well,
\begin{equation}
\lim_{K\rightarrow\infty}\sum_{k=0}^{K}\beta_{k}^{2}\left\Vert \widehat{\boldsymbol{g}}_{k}\right\Vert ^{2}<\infty.\label{eq:c3}
\end{equation}
From the above equations (\ref{eq:c1})-(\ref{eq:c3}), we conclude
that there exists $W<\infty$ such that $d_{K+1}\leq W+z_{K}$, with
\begin{equation}
z_{K}=2\alpha_{2}\sum_{k=0}^{K}\beta_{k}\gamma_{k}\left(\boldsymbol{a}_{k}-\boldsymbol{a}^{*}\right)^{T}\cdot\left(\nabla F\left(\boldsymbol{a}_{k}\right)+\boldsymbol{b}_{k}\right).\label{eq:c13}
\end{equation}
Since $d_{K+1}\geq0$ by definition, we have $z_{K}>-\infty$. 

Recheck (\ref{eq:bound_d}), we can say that for an arbitrary small
positive value $\varepsilon_{b}$, there exists $K_{b}$ such that,
\begin{equation}
\left\Vert \nabla F\left(\boldsymbol{a}_{k}\right)+\boldsymbol{b}_{k}\right\Vert \geq\left(1-\varepsilon_{b}\right)\left\Vert \nabla F\left(\boldsymbol{a}_{k}\right)\right\Vert ,\:\forall k\geq K_{b},\label{eq:c4}
\end{equation}
which also implies that $\left(\boldsymbol{a}_{k}-\boldsymbol{a}^{*}\right)^{T}\cdot\left(\nabla F\left(\boldsymbol{a}_{k}\right)+\boldsymbol{b}_{k}\right)\leq0$,
$\forall k\geq K_{b}$, by the concavity of $F$ in (\ref{eq:concave}).
Therefore $0\leq d_{K+1}<\infty$ for any large $K$ and $\lim_{K\rightarrow\infty}d_{K+1}=\overline{d}$
exists. The following steps of the proof is similar to the classical
proof in \cite{robbins1951stochastic}. 

Assume that: \emph{H1}) $\overline{d}>0$, \emph{i.e.}, $\boldsymbol{a}_{k}$
does not converge to $\boldsymbol{a}^{*}$, then for any $\varepsilon_{h}>0$,
there exists $K_{h}$ such that 
\begin{equation}
\left(\boldsymbol{a}_{k}-\boldsymbol{a}^{*}\right)^{T}\cdot\nabla F\left(\boldsymbol{a}_{k}\right)<-\varepsilon_{h},\quad\forall k\geq K_{h}.\label{eq:ch}
\end{equation}
From (\ref{eq:c13}) and (\ref{eq:c4}), we get that $\forall k\geq K_{m}=\max\left\{ K_{h},K_{b}\right\} $,
\begin{equation}
\left(\boldsymbol{a}_{k}-\boldsymbol{a}^{*}\right)^{T}\cdot\left(\nabla F\left(\boldsymbol{a}_{k}\right)+\boldsymbol{b}_{k}\right)<-\varepsilon_{h}\left(1-\varepsilon_{b}\right),\label{eq:cbh}
\end{equation}
which leads to 
\begin{align}
 & \lim_{K\rightarrow\infty}\sum_{k=K_{m}}^{K}\beta_{k}\gamma_{k}\left(\boldsymbol{a}_{k}-\boldsymbol{a}^{*}\right)^{T}\cdot\left(\nabla F\left(\boldsymbol{a}_{k}\right)+\boldsymbol{b}_{k}\right)\nonumber \\
 & <-\varepsilon_{h}\left(1-\varepsilon_{b}\right)\lim_{K\rightarrow\infty}\sum_{k=K_{m}}^{K}\beta_{k}\gamma_{k}<-\infty,
\end{align}
as $\sum\beta_{k}\gamma_{k}$ diverges by assumption \emph{A4}. 

We find that $z_{K}<-\infty$ and $d_{K+1}<-\infty$. However $d_{K+1}$
should be positive by definition. Therefore, the hypothesis \emph{H1}
cannot be true, there should be $\lim_{k\rightarrow\infty}d_{k}=0$,
$\lim_{k\rightarrow\infty}\nabla F\left(\boldsymbol{a}_{k}\right)=\mathbf{0}$,
and $\lim_{k\rightarrow\infty}\boldsymbol{a}_{k}=\boldsymbol{a}^{*}$
almost surely, which concludes the proof.

\subsection{\label{subsec:Proof-constrained}Proof of Theorem~\ref{prop:converge_constraint}}

We learn first the property of the projection (\ref{eq:proj_sp2})
in order to find an efficient way to simplify the proof. Otherwise,
as stated in Remark~\ref{rem_const}, the proof can be complicated. 

Define $\mathcal{C}_{i,k}=\left[a_{i,\min}+\alpha_{3}\gamma_{k},a_{i,\max}-\alpha_{3}\gamma_{k}\right]$
for any $i\in\mathcal{N}$ and $\mathcal{C}_{k}=\mathcal{C}_{1,k}\times\ldots\times\mathcal{C}_{N,k}$.
Obviously, $\mathtt{Proj}\left(\boldsymbol{a}_{k}+\beta_{k}\widehat{\boldsymbol{g}}_{k}\right)\in\mathcal{C}_{k+1}$
by (\ref{eq:proj_sp2}). Since $\gamma_{k}$ is a decreasing sequence,
we find that $\mathcal{C}_{k}\subseteq\mathcal{C}_{k+1}$. Furthermore,
we have $\lim_{k\rightarrow\infty}\mathcal{C}_{k}=\left[0,a_{\max}\right]^{N}$
as $\lim_{k\rightarrow\infty}\gamma_{k}=0$. Hence, there exists $K_{\mathrm{c}}\in\mathbb{N}$,
such that $\boldsymbol{a}^{*}\in\mathcal{C}_{k+1}$ for any $k\geq K_{\mathrm{c}}$.
It is straightforward to prove that 
\[
\left\Vert \mathtt{Proj}\left(\boldsymbol{a}_{k}+\beta_{k}\widehat{\boldsymbol{g}}_{k}\right)-\boldsymbol{a}^{*}\right\Vert \leq\left\Vert \boldsymbol{a}_{k}+\beta_{k}\widehat{\boldsymbol{g}}_{k}-\boldsymbol{a}^{*}\right\Vert ,\:\forall k\geq K_{\mathrm{c}},
\]
meaning that the projection makes $\boldsymbol{a}_{k+1}$ closer to
$\boldsymbol{a}^{*}$ in the Euclidean space when $\boldsymbol{a}_{k+1}\in\mathcal{C}_{k+1}$
and $\boldsymbol{a}^{*}\in\mathcal{C}_{k+1}$.

Re-consider the divergence as defined in (\ref{eq:divergence_def}),
we have
\begin{align}
d_{k+1} & =\left\Vert \mathtt{Proj}\left(\boldsymbol{a}_{k}+\beta_{k}\widehat{\boldsymbol{g}}_{k}\right)-\boldsymbol{a}^{*}\right\Vert ^{2}.\nonumber \\
 & \leq\left\Vert \boldsymbol{a}_{k}+\beta_{k}\widehat{\boldsymbol{g}}_{k}-\boldsymbol{a}^{*}\right\Vert ^{2},\quad\forall k\geq K_{\mathrm{c}}.\label{eq:divergence_def_1}
\end{align}
from which we get
\begin{align}
 & d_{K+1}\leq d_{K_{\mathrm{c}}}+\sum_{k=K_{\mathrm{c}}}^{K}\beta_{k}^{2}\left\Vert \widehat{\boldsymbol{g}}_{k}\right\Vert ^{2}+2\sum_{k=K_{\mathrm{c}}}^{K}\beta_{k}\left(\boldsymbol{a}_{k}-\boldsymbol{a}^{*}\right)^{T}\cdot\boldsymbol{e}_{k}\nonumber \\
 & \qquad+2\alpha_{2}\sum_{k=K_{\mathrm{c}}}^{K}\beta_{k}\gamma_{k}\left(\boldsymbol{a}_{k}-\boldsymbol{a}^{*}\right)^{T}\cdot\left(\nabla F\left(\boldsymbol{a}_{k}\right)+\boldsymbol{b}_{k}\right),\label{eq:c1-1}
\end{align}
where $\boldsymbol{b}_{k}$ and $\boldsymbol{e}_{k}$ are the same
as those defined in (\ref{eq:bias_def}) and (\ref{eq:sto_n_def}).
The results presented in Lemma~\ref{lem:bias} and in Lemma~\ref{lem:stochastic_noise}
are valid as well. 

Clearly, (\ref{eq:c1-1}) has similar shape compared with (\ref{eq:c1}).
Using similar steps in Appendix~\ref{subsec:proof_conv}, Theorem~\ref{prop:converge_constraint}
can be proved.

\subsection{Proof of Lemma \ref{lem:aver_f_in}}

\label{subsec:Proof_in}We start with the conditional expectation,
based on (\ref{eq:f_approx-1}), we have 
\begin{align}
 & \mathbb{E}_{\mathcal{I}_{i,k}}\left(\widetilde{f}_{i}^{(\textrm{I})}\left(\boldsymbol{a},\mathbf{S}_{k},\mathcal{I}_{i,k}\right)\big|\left|\mathcal{I}_{i,k}\right|=n\right)\nonumber \\
 & =\begin{cases}
\widetilde{u}_{i,k}+\frac{N-1}{n}\mathbb{E}_{\mathcal{I}_{i,k}}\left(\sum_{j\in\mathcal{I}_{i,k}}\widetilde{u}_{j,k}\big|\left|\mathcal{I}_{i,k}\right|=n\right), & \textrm{if }n\neq0,\\
0, & \textrm{if }n=0.
\end{cases}\label{eq:condindex}
\end{align}
Denote $\mathcal{U}_{i}^{\left(n\right)}$ as a collection of all
possible sets $\mathcal{I}_{i,k}$ such that $\left|\mathcal{I}_{i,k}\right|=n$,
\emph{e.g.}, $\mathcal{U}_{i}^{\left(1\right)}=\left\{ \left\{ 1\right\} ,\ldots,\left\{ i-1\right\} ,\left\{ i+1\right\} ,\ldots\right\} $.
Since each node has an equal probability to be involved in $\mathcal{I}_{i,k}$,
the sets in $\mathcal{U}_{i}^{\left(n\right)}$ are also equiprobable,
\emph{i.e.}, 
\[
\mathbb{P}\left(\mathcal{I}_{i,k}=\mathcal{I}\big|\left|\mathcal{I}_{i,k}\right|=n\right)=\frac{1}{\binom{N-1}{n}},\quad\forall\mathcal{I}\in\mathcal{U}_{i}^{\left(n\right)},
\]
note that the cardinal of $\mathcal{U}_{i}^{\left(n\right)}$ is $\binom{N-1}{n}$.
We evaluate
\begin{align}
 & \mathbb{E}_{\mathcal{I}_{i,k}}\left(\sum_{j\in\mathcal{I}_{i,k}}\widetilde{u}_{j,k}\bigg|\left|\mathcal{I}_{i,k}\right|=n\right)\nonumber \\
 & =\sum_{\mathcal{I}\in\mathcal{U}_{i}^{\left(n\right)}}\frac{1}{\binom{N-1}{n}}\sum_{j\in\mathcal{I}}\widetilde{u}_{j,k}=\frac{1}{\binom{N-1}{n}}\frac{n\binom{N-1}{n}}{N-1}\sum_{j\in\mathcal{N}\setminus\left\{ i\right\} }\widetilde{u}_{j,k}\nonumber \\
 & =\frac{n}{N-1}\sum_{j\in\mathcal{N}\setminus\left\{ i\right\} }\widetilde{u}_{j,k}.\label{eq:condindex_2}
\end{align}
Combine (\ref{eq:condindex}) and (\ref{eq:condindex_2}), for any
$n\in\left\{ 1,\ldots,N-1\right\} $, we have
\begin{align}
\mathbb{E}_{\mathcal{I}_{i,k}}\left(\widetilde{f}_{i}^{(\textrm{I})}\left(\boldsymbol{a},\mathbf{S}_{k},\mathcal{I}_{i,k}\right)\big|\left|\mathcal{I}_{i,k}\right|=n\right) & =\widetilde{f}\left(\boldsymbol{a},\mathbf{S}_{k}\right).\label{eq:condindex-1}
\end{align}
According to the basic rule of expectation,
\begin{align*}
 & \mathbb{E}\left(\widetilde{f}_{i}^{(\textrm{I})}\left(\boldsymbol{a},\mathbf{S}_{k},\mathcal{I}_{i,k}\right)\right)\\
 & =\sum_{n=0}^{N-1}\mathbb{P}\left(\left|\mathcal{I}_{i,k}\right|=n\right)\mathbb{E}_{\mathcal{I}_{i,k}}\left(\widetilde{f}_{i}^{(\textrm{I})}\left(\boldsymbol{a},\mathbf{S}_{k},\mathcal{I}_{i,k}\right)\big|\left|\mathcal{I}_{i,k}\right|=n\right)\\
 & =\sum_{n=1}^{N-1}\mathbb{P}\left(\left|\mathcal{I}_{i,k}\right|=n\right)\widetilde{f}\left(\boldsymbol{a},\mathbf{S}_{k}\right)\\
 & =\left(1-\mathbb{P}\left(\left|\mathcal{I}_{i,k}\right|=0\right)\right)\widetilde{f}\left(\boldsymbol{a},\mathbf{S}_{k}\right)\\
 & =\left(1-(1-p)^{N}\right)\widetilde{f}\left(\boldsymbol{a},\mathbf{S}_{k}\right),
\end{align*}
which concludes the proof. 

\subsection{Proof of Theorem~\ref{prop:converge_rp_in}}

\label{subsec:Proof-of-convergence_in}As the analysis in Section
\ref{subsec:Convergence-results}, we mainly need to check: 
\begin{itemize}
\item \emph{C1}) whether $b_{i,k}^{\left(\textrm{I}\right)}$ is vanishing;
\item \emph{C2}) whether $\sum_{k=1}^{\infty}\beta_{k}\boldsymbol{e}_{k}^{\left(\textrm{I}\right)}$
converges.
\end{itemize}
The proof of these two statements are more complicated as compared
to Appendix~\ref{subsec:Proof-bias} and Appendix~\ref{subsec:Proof-e},
since there is an additional random term $\mathcal{I}_{i,k}$ in both
$b_{i,k}^{\left(\textrm{I}\right)}$ and $e_{i,k}^{\left(\textrm{I}\right)}$,
compared with $b_{i,k}$ and $e_{i,k}$ discussed in Section~\ref{sec:Distributed-allocation-algorithm}.
Then Theorem~\ref{prop:converge_rp_in} can be proved in the end.

\subsubsection{Proof of C1}

We start with the bias term, by Lemma~\ref{lem:aver_f_in}, we have
\begin{align*}
\mathbb{E}_{\mathbf{S},\boldsymbol{\Phi},\boldsymbol{\eta},\mathcal{I}_{i,k}}\left(\widehat{g}_{i,k}^{(\textrm{I})}\right) & =\mathbb{E}_{\mathbf{S},\boldsymbol{\Phi},\boldsymbol{\eta},\mathcal{I}_{i,k}}\left(\Phi_{i,k}\widetilde{f}_{i}^{(\textrm{I})}\left(\boldsymbol{a},\mathbf{S}_{k},\mathcal{I}_{i,k}\right)\right)\\
 & =q\mathbb{E}_{\mathbf{S},\boldsymbol{\Phi},\boldsymbol{\eta}}\left(\Phi_{i,k}\widetilde{f}\left(\boldsymbol{a},\mathbf{S}_{k}\right)\right).
\end{align*}
We then find that $b_{i,k}^{\left(\textrm{I}\right)}$ in this section
is the same as $b_{i,k}$ defined in (\ref{eq:bias_def}), as
\begin{align}
b_{i,k}^{\left(\textrm{I}\right)} & =\frac{q\mathbb{E}_{\mathbf{S},\boldsymbol{\Phi},\boldsymbol{\eta}}\left(\Phi_{i,k}\widetilde{f}\left(\boldsymbol{a},\mathbf{S}_{k}\right)\right)}{q\alpha_{2}\gamma_{k}}-F_{i}'\left(\boldsymbol{a}_{k}\right)\nonumber \\
 & =b_{i,k}.\label{eq:b_I}
\end{align}
Therefore, $b_{i,k}^{\left(\textrm{I}\right)}$ can be vanishing as
$\gamma_{k}$ is vanishing according to Lemma~\ref{lem:bias}, \emph{C1}
is verified.

\subsubsection{Proof of C2}

The second step is to analyze the stochastic noise to check \emph{C2}.
From the definition (\ref{eq:sto_n_in}) , we have
\begin{align*}
\mathbb{E}_{\mathbf{S},\boldsymbol{\Phi},\boldsymbol{\eta},\mathcal{I}_{i,k}}\left(e_{i,k}^{\left(\textrm{I}\right)}\right) & =\mathbb{E}_{\mathbf{S},\boldsymbol{\Phi},\boldsymbol{\eta},\mathcal{I}_{i,k}}\left(\widehat{g}_{i,k}^{\left(\textrm{I}\right)}-\mathbb{E}_{\mathbf{S},\boldsymbol{\Phi},\boldsymbol{\eta},\mathcal{I}_{i,k}}\left(\widehat{g}_{i,k}^{\left(\textrm{I}\right)}\right)\right)\\
 & =0.
\end{align*}
As $\mathcal{I}_{i,k}$ and $\mathcal{I}_{i,k'}$ are independent
for any $k\neq k'$, the sequence $\left\{ \sum_{k=K}^{K'}\gamma_{k}\boldsymbol{e}_{k}^{\left(\textrm{I}\right)}\right\} _{K'\geq K}$
is martingale. By applying Doob's inequality, we have
\begin{align}
 & \mathbb{P}\left(\sup_{K'\geq K}\left\Vert \sum_{k=K}^{K'}\beta_{k}\boldsymbol{e}_{k}^{\left(\textrm{I}\right)}\right\Vert \geq\rho\right)\nonumber \\
 & \leq\frac{1}{\rho^{2}}\mathbb{E}_{\mathbf{S},\boldsymbol{\Phi},\boldsymbol{\eta},\mathcal{I}_{i,k}}\left(\left\Vert \sum_{k=K}^{K'}\beta_{k}\boldsymbol{e}_{k}\right\Vert ^{2}\right)\nonumber \\
 & \overset{\left(a\right)}{\leq}\frac{1}{\rho^{2}}\sum_{k=K}^{\infty}\beta_{k}^{2}\sum_{i\in\mathcal{N}}\mathbb{E}_{\mathbf{S},\boldsymbol{\Phi},\boldsymbol{\eta},\mathcal{I}_{i,k}}\left(\left(\widehat{g}_{i,k}^{\left(\textrm{I}\right)}\right)^{2}\right),\label{eq:e_up_1-1}
\end{align}
where $\left(a\right)$ can be obtained using similar steps in (\ref{eq:e_up_1}).
In order to evaluate the average of $\left(\widehat{g}_{i,k}^{\left(\textrm{I}\right)}\right)^{2}$,
we need to consider, for any $n\in\left\{ 1,\ldots,N-1\right\} $,
\begin{align}
 & \mathbb{E}_{\mathcal{I}_{i,k}}\left(\left(\widehat{g}_{i,k}^{\left(\textrm{I}\right)}\right)^{2}\big|\left|\mathcal{I}_{i,k}\right|=n\right)\nonumber \\
 & =\Phi_{i,k}^{2}\mathbb{E}_{\mathcal{I}_{i,k}}\left(\left.\left(\widetilde{u}_{i,k}+\frac{N-1}{n}\sum_{j\in\mathcal{I}_{i,k}}\widetilde{u}_{j,k}\right)^{2}\right|\left|\mathcal{I}_{i,k}\right|=n\right)\nonumber \\
 & \overset{(a)}{\leq}\alpha_{3}^{2}\left(n+1\right)\left(\widetilde{u}_{i,k}^{2}+\left(\frac{N-1}{n}\right)^{2}\cdot\right.\nonumber \\
 & \qquad\qquad\qquad\qquad\left.\mathbb{E}_{\mathcal{I}_{i,k}}\left(\left.\sum_{j\in\mathcal{I}_{i,k}}\widetilde{u}_{j,k}^{2}\right|\left|\mathcal{I}_{i,k}\right|=n\right)\right)\nonumber \\
 & \overset{(b)}{=}\alpha_{3}^{2}\left(n+1\right)\left(\widetilde{u}_{i,k}^{2}+\frac{N-1}{n}\sum_{j\in\mathcal{N}\setminus\left\{ i\right\} }\widetilde{u}_{j,k}^{2}\right)\nonumber \\
 & \overset{(c)}{\leq}\alpha_{3}^{2}\left(\left(n+1\right)\widetilde{u}_{i,k}^{2}+2\left(N-1\right)\sum_{j\in\mathcal{N}\setminus\left\{ i\right\} }\widetilde{u}_{j,k}^{2}\right),\label{eq:e_up_in_2}
\end{align}
where $(a)$ is by $\Phi_{i,k}^{2}\leq\alpha_{3}^{2}$ and $\sum_{i=1}^{m}x_{i}/m\leq\sqrt{\sum_{i=1}^{m}x_{i}^{2}/m}$,
$(b)$ can be proved using the same way as (\ref{eq:condindex_2}),
and $(c)$ is due to $1<\left(n+1\right)/n\leq2$ as $1\leq n\leq N-1$. 

Since $\widehat{g}_{i,k}^{\left(\textrm{I}\right)}=0$ as $n=0$ by
definition, we have 
\begin{align}
 & \mathbb{E}_{\mathcal{I}_{i,k}}\left(\left(\widehat{g}_{i,k}^{\left(\textrm{I}\right)}\right)^{2}\right)\nonumber \\
 & =\sum_{n=1}^{N-1}\mathbb{P}\left(\left|\mathcal{I}_{i,k}\right|=n\right)\mathbb{E}_{\mathcal{I}_{i,k}}\left(\left.\left(\widehat{g}_{i,k}^{\left(\textrm{I}\right)}\right)^{2}\right|\left|\mathcal{I}_{i,k}\right|=n\right)\nonumber \\
 & \leq\mathbb{E}_{\mathcal{I}_{i,k}}\left(\left|\mathcal{I}_{i,k}\right|\right)\alpha_{3}^{2}\widetilde{u}_{i,k}^{2}+\nonumber \\
 & \qquad\mathbb{P}\left(\left|\mathcal{I}_{i,k}\right|\neq0\right)\cdot\alpha_{3}^{2}\left(\widetilde{u}_{i,k}^{2}+2\left(N-1\right)\sum_{j\in\mathcal{N}\setminus\left\{ i\right\} }\widetilde{u}_{j,k}^{2}\right)\nonumber \\
 & =\alpha_{3}^{2}\left(\left(p\left(N-1\right)+q\right)\widetilde{u}_{i,k}^{2}+2\left(N-1\right)q\sum_{j\in\mathcal{N}\setminus\left\{ i\right\} }\widetilde{u}_{j,k}^{2}\right),\label{eq:g_s_set}
\end{align}
where $\mathbb{E}_{\mathcal{I}_{i,k}}\left(\left|\mathcal{I}_{i,k}\right|\right)=\left(N-1\right)p$
and $q=\mathbb{P}\left(\left|\mathcal{I}_{i,k}\right|\neq0\right)$
is defined in (\ref{eq:q_def}). 

We also need to find a upper bound of $\mathbb{E}_{\mathbf{S},\boldsymbol{\Phi},\boldsymbol{\eta}}\left(\widetilde{u}_{i,k}^{2}\right)$
for any $i\in\mathcal{N}$, the following steps are similar to (\ref{eq:up_2}),
\begin{align}
 & \mathbb{E}_{\mathbf{S},\boldsymbol{\Phi},\boldsymbol{\eta}}\left(\widetilde{u}_{i,k}^{2}\right)\nonumber \\
 & =\mathbb{E}_{\mathbf{S},\boldsymbol{\Phi},\boldsymbol{\eta}}\left(\left(u_{i}\left(\boldsymbol{a}_{k}+\gamma_{k}\boldsymbol{\Phi}_{k},\mathbf{S}_{k}\right)+\eta_{i,k}\right)^{2}\right)\nonumber \\
 & =\mathbb{E}_{\mathbf{S},\boldsymbol{\Phi}}\left(\left(u_{i}\left(\boldsymbol{a}_{k}+\gamma_{k}\boldsymbol{\Phi}_{k},\mathbf{S}_{k}\right)\right)^{2}\right)+\alpha_{4}\nonumber \\
 & \overset{(a)}{\leq}\mathbb{E}_{\mathbf{S},\boldsymbol{\Phi}}\left(\left(\left\Vert u_{i}\left(\mathbf{0},\mathbf{S}_{k}\right)\right\Vert +L_{\mathbf{S}_{k}}\left\Vert \boldsymbol{a}_{k}+\gamma_{k}\boldsymbol{\Phi}_{k}\right\Vert \right)^{2}\right)+\alpha_{4}\nonumber \\
 & \overset{(b)}{\leq}2\mathbb{E}_{\mathbf{S}}\left(\iota_{\mathbf{S}_{k}}^{2}+L_{\mathbf{S}_{k}}^{2}\left(\left\Vert \boldsymbol{a}_{k}\right\Vert +\sqrt{N}\gamma_{k}\alpha_{3}\right)^{2}\right)+\alpha_{4}\nonumber \\
 & \overset{(c)}{=}2\left(\iota+L\left(\left\Vert \boldsymbol{a}_{k}\right\Vert +\sqrt{N}\gamma_{k}\alpha_{3}\right)^{2}\right)+\alpha_{4}<\infty\label{eq:up_2-1}
\end{align}
where: $(a)$ is by (\ref{eq:Lipschitz_f}); in $(b)$, we introduce
$\iota_{\mathbf{S}_{k}}=\max_{i\in\mathcal{N}}\left\{ \left\Vert u_{i}\left(\mathbf{0},\mathbf{S}_{k}\right)\right\Vert \right\} <\infty$
as an bounded upper bound of $\left\Vert u_{i}\left(\mathbf{0},\mathbf{S}_{k}\right)\right\Vert $;
in $(c)$, we denote $\iota=\mathbb{E}_{\mathbf{S}}\left(\iota_{\mathbf{S}_{k}}^{2}\right)$
and $L=\mathbb{E}_{\mathbf{S}}\left(L_{\mathbf{S}_{k}}^{2}\right)$.
Note that the upper bound (\ref{eq:up_2-1}) is valid for any $i\in\mathcal{N}$.

Based on (\ref{eq:g_s_set}) and (\ref{eq:up_2-1}), we evaluate
\begin{align}
 & \mathbb{E}_{\mathbf{S},\boldsymbol{\Phi},\boldsymbol{\eta},\mathcal{I}_{i,k}}\left(\left(\widehat{g}_{i,k}^{\left(\textrm{I}\right)}\right)^{2}\right)\nonumber \\
 & =\mathbb{E}_{\mathbf{S},\boldsymbol{\Phi},\boldsymbol{\eta}}\left(\mathbb{E}_{\mathcal{I}_{i,k}}\left(\left(\widehat{g}_{i,k}^{\left(\textrm{I}\right)}\right)^{2}\right)\right)\nonumber \\
 & \leq\alpha_{3}^{2}\left(2\left(N-1\right)^{2}q+\left(N-1\right)p+q\right)\cdot\nonumber \\
 & \qquad\qquad\left(2\left(\iota+L\left(\left\Vert \boldsymbol{a}_{k}\right\Vert +\sqrt{N}\gamma_{k}\alpha_{3}\right)^{2}\right)+\alpha_{4}\right)\nonumber \\
 & <\infty.\label{eq:bound_g_i}
\end{align}

From (\ref{eq:e_up_1-1}) and (\ref{eq:bound_g_i}), we can conclude
that $\sum_{k=1}^{\infty}\beta_{k}\boldsymbol{e}_{k}^{\left(\textrm{I}\right)}$
converges almost surely. 

\subsubsection{Proof of convergence }

Finally, we can show the convergence of Algorithm~\ref{alg:ESSP-based-Algorithm-1}.
Consider the evolution of the divergence $d_{k}^{\left(\mathrm{I}\right)}=\left\Vert \boldsymbol{a}_{k}^{\left(\mathrm{I}\right)}-\boldsymbol{a}^{*}\right\Vert ^{2}$,
we get 
\begin{align}
d_{K+1}^{\left(\mathrm{I}\right)} & =d_{0}+\sum_{k=0}^{K}\beta_{k}^{2}\left\Vert \widehat{\boldsymbol{g}}_{k}^{\left(\mathrm{I}\right)}\right\Vert ^{2}+2\sum_{k=0}^{K}\beta_{k}\left(\boldsymbol{a}_{k}^{\left(\mathrm{I}\right)}-\boldsymbol{a}^{*}\right)^{T}\cdot\boldsymbol{e}_{k}^{\left(\mathrm{I}\right)}\nonumber \\
 & +2\alpha_{2}\sum_{k=0}^{K}\beta_{k}\gamma_{k}\left(\boldsymbol{a}_{k}^{\left(\mathrm{I}\right)}-\boldsymbol{a}^{*}\right)^{T}\cdot\left(\nabla F\left(\boldsymbol{a}_{k}^{\left(\mathrm{I}\right)}\right)+\boldsymbol{b}_{k}^{\left(\mathrm{I}\right)}\right),\label{eq:c1-2}
\end{align}
with the same shape as (\ref{eq:c1}) and be obtained using the same
steps. The following proof is also similar to that in Appendix~\ref{subsec:proof_conv},
as we have already shown that $\boldsymbol{b}_{k}^{\left(\mathrm{I}\right)}$
and $\boldsymbol{e}_{k}^{\left(\mathrm{I}\right)}$ have similar properties
as $\boldsymbol{b}_{k}$ and $\boldsymbol{e}_{k}$.

\subsection{\label{subsec:Proof-of-Lemma_D}Proof of Lemma \ref{lem:inequality_D}}

We investigate the relation between the two successive average divergence.
We mainly demonstrate (\ref{eq:dk+1_dk}), as the proof of (\ref{eq:dk+1_dk_I})
is quite similar. 

In Algorithm~\ref{alg:ESSP-based-Algorithm}, we have 
\begin{align}
D_{k+1}^{\left(\mathrm{C}\right)} & =\mathbb{E}\left(\left\Vert \boldsymbol{a}_{k+1}-\boldsymbol{a}^{*}\right\Vert _{2}^{2}\right)=\mathbb{E}\left(\left\Vert \boldsymbol{a}_{k}+\beta_{k}\widehat{\boldsymbol{g}}_{k}-\boldsymbol{a}^{*}\right\Vert _{2}^{2}\right)\nonumber \\
 & =D_{k}^{\left(\mathrm{C}\right)}+\beta_{k}^{2}\mathbb{E}\left(\left\Vert \widehat{\boldsymbol{g}}_{k}\right\Vert _{2}^{2}\right)+2\beta_{k}\mathbb{E}\left(\left(\boldsymbol{a}_{k}-\boldsymbol{a}^{*}\right)^{T}\widehat{\boldsymbol{g}}_{k}\right).\label{eq:Dk+1_Dk}
\end{align}
According to (\ref{eq:up_2}), we have shown that $\mathbb{E}\left(\left\Vert \widehat{\boldsymbol{g}}_{k}\right\Vert _{2}^{2}\right)<M^{\left(\mathrm{C}\right)}$
almost surely, with $M^{\left(\mathrm{C}\right)}$ a bounded constant.
Meanwhile, we evaluate
\begin{align}
\mathbb{E}\left(\left(\boldsymbol{a}_{k}-\boldsymbol{a}^{*}\right)^{T}\widehat{\boldsymbol{g}}_{k}\right) & =\alpha_{2}\gamma_{k}\mathbb{E}\left(\left(\boldsymbol{a}_{k}-\boldsymbol{a}^{*}\right)^{T}\left(\nabla F\left(\boldsymbol{a}_{k}\right)+\boldsymbol{b}_{k}\right)\right),\label{eq:order1_0}
\end{align}
which can be obtained by using (\ref{eq:RM_form}) and $\mathbb{E}\left(\boldsymbol{e}_{k}\right)=\mathbf{0}$.
We have 
\begin{align}
\left(\boldsymbol{a}_{k}-\boldsymbol{a}^{*}\right)^{T}\boldsymbol{b}_{k} & \leq\sum_{i=1}^{N}\left|a_{i,k}-a_{i}^{*}\right|\left|b_{i,k}\right|\nonumber \\
 & \leq N^{2}\frac{\alpha_{3}^{3}\alpha_{1}}{2\alpha_{2}}\gamma_{k}\sum_{i=1}^{N}\left|a_{i,k}-a_{i}^{*}\right|\nonumber \\
 & \leq N^{2}\frac{\alpha_{3}^{3}\alpha_{1}}{2\alpha_{2}}\gamma_{k}\sqrt{N\sum_{i=1}^{N}\left(a_{i,k}-a_{i}^{*}\right)^{2}}\nonumber \\
 & =N^{\frac{5}{2}}\frac{\alpha_{3}^{3}\alpha_{1}}{2\alpha_{2}}\gamma_{k}\left\Vert \boldsymbol{a}_{k}-\boldsymbol{a}^{*}\right\Vert _{2}.\label{eq:order1_1}
\end{align}

From (\ref{eq:order1_0}), (\ref{eq:order1_1}), and the strong concavity
(\ref{eq:order1_2}), we have 
\begin{align}
 & \mathbb{E}\left(\left(\boldsymbol{a}_{k}-\boldsymbol{a}^{*}\right)^{T}\widehat{\boldsymbol{g}}_{k}\right)\nonumber \\
 & \leq\alpha_{2}\gamma_{k}\mathbb{E}\left(-\alpha_{5}\left\Vert \boldsymbol{a}_{k}-\boldsymbol{a}^{*}\right\Vert _{2}^{2}+N^{\frac{5}{2}}\frac{\alpha_{3}^{3}\alpha_{1}}{2\alpha_{2}}\gamma_{k}\left\Vert \boldsymbol{a}_{k}-\boldsymbol{a}^{*}\right\Vert _{2}\right)\nonumber \\
 & =-\alpha_{2}\alpha_{5}\gamma_{k}\mathbb{E}\left(\left\Vert \boldsymbol{a}_{k}-\boldsymbol{a}^{*}\right\Vert _{2}^{2}\right)\nonumber \\
 & \qquad\qquad\qquad\qquad+N^{\frac{5}{2}}\frac{\alpha_{3}^{3}\alpha_{1}}{2}\gamma_{k}^{2}\sqrt{\left|\mathbb{E}\left(\left\Vert \boldsymbol{a}_{k}-\boldsymbol{a}^{*}\right\Vert _{2}\right)\right|^{2}}\nonumber \\
 & =-\alpha_{2}\alpha_{5}\gamma_{k}D_{k}+\frac{1}{2}N^{\frac{5}{2}}\alpha_{3}^{3}\alpha_{1}\gamma_{k}^{2}\sqrt{D_{k}}.\label{eq:order1_}
\end{align}

By combining (\ref{eq:Dk+1_Dk}), (\ref{eq:order1_}) and the fact
that $\mathbb{E}\left(\left\Vert \widehat{\boldsymbol{g}}_{k}\right\Vert _{2}^{2}\right)\leq M^{\left(\mathrm{C}\right)}$,
we can obtain (\ref{eq:dk+1_dk}).

We have the similar steps to prove (\ref{eq:dk+1_dk_I}) for Algorithm~\ref{alg:ESSP-based-Algorithm-1},
the main difference comes from the upper bound of $\mathbb{E}(\left\Vert \widehat{\boldsymbol{g}}_{k}^{\left(\mathrm{I}\right)}\right\Vert _{2}^{2})$
and 
\begin{align}
 & \mathbb{E}\left(\left(\boldsymbol{a}_{k}-\boldsymbol{a}^{*}\right)^{T}\widehat{\boldsymbol{g}}_{k}^{\left(\mathrm{I}\right)}\right)\nonumber \\
 & =q\alpha_{2}\gamma_{k}\mathbb{E}\left(\left(\boldsymbol{a}_{k}-\boldsymbol{a}^{*}\right)^{T}\left(\nabla F\left(\boldsymbol{a}_{k}\right)+\boldsymbol{b}_{k}^{\left(\mathrm{I}\right)}\right)\right)\nonumber \\
 & \leq q\left(-\alpha_{2}\alpha_{5}\gamma_{k}D_{k}+\frac{1}{2}N^{\frac{5}{2}}\alpha_{3}^{3}\alpha_{1}\gamma_{k}^{2}\sqrt{D_{k}}\right),
\end{align}
which is similar to (\ref{eq:order1_}) and the difference comes from
the presence of the parameter $q$ in the generalized Robbins-Monro
form (\ref{eq:RM_form-1}), note that $\boldsymbol{b}_{k}^{\left(\mathrm{I}\right)}=\boldsymbol{b}_{k}$
as shown in (\ref{eq:b_I}).

\subsection{\label{subsec:Proof-necerate}Proof of Lemma~\ref{lem:nece_rate}}

From (\ref{eq:dk+1_dk_0}) and $D_{k}\leq U_{k}$, we get
\begin{align*}
D_{k+1} & \leq\left(1-A\beta_{k}\gamma_{k}\right)U_{k}+B\beta_{k}\gamma_{k}^{2}\sqrt{U_{k}}+C\beta_{k}^{2},
\end{align*}
because $1-A\beta_{k}\gamma_{k}>0$ when $k\geq K_{0}$. In order
to perform the induction, we need to have 
\[
\left(1-A\beta_{k}\gamma_{k}\right)U_{k}+B\beta_{k}\gamma_{k}^{2}\sqrt{U_{k}}+C\beta_{k}^{2}\leq U_{k+1}\leq U_{k},
\]
which leads to
\begin{equation}
A\gamma_{k}U_{k}-B\gamma_{k}^{2}\sqrt{U_{k}}-C\beta_{k}\geq0.\label{eq:necessary_rate}
\end{equation}
By solving (\ref{eq:necessary_rate}), we obtain (\ref{eq:lower_u})
as $\sqrt{U_{k}}>0$. 

\subsection{Proof of Theorem \ref{prop:rate}}

\label{subsec:proof_convergencerate}

We start with the proof of (\ref{eq:Dk_up1}), which is realized by
induction.

It is straightforward to get $D_{K_{0}}\leq\vartheta^{2}\gamma_{K_{0}}^{2}$
according to the definition of $\vartheta$. We mainly needs to verify
that $D_{k}\leq\vartheta^{2}\gamma_{k}^{2}$ leads to $D_{k+1}\leq\vartheta^{2}\gamma_{k+1}^{2}$
for any $k\geq K_{0}$. 

Suppose that $D_{k}\leq\vartheta^{2}\gamma_{k}^{2}$, from (\ref{eq:dk+1_dk_0}),
we have 
\begin{align*}
D_{k+1} & \leq\left(1-A\beta_{k}\gamma_{k}\right)\gamma_{k}^{2}\vartheta^{2}+B\beta_{k}\gamma_{k}^{3}\vartheta+C\beta_{k}^{2}.
\end{align*}
We then need to show that there exists $\vartheta\in\mathbb{R}^{+}$,
such that
\begin{align*}
\left(1-A\beta_{k}\gamma_{k}\right)\gamma_{k}^{2}\vartheta^{2}+B\beta_{k}\gamma_{k}^{3}\vartheta+C\beta_{k}^{2} & \leq U_{k+1}\\
 & =\vartheta^{2}\gamma_{k+1}^{2},
\end{align*}
which can be written as
\begin{equation}
\left(\chi_{k}-A\right)\vartheta^{2}+B\vartheta+C\beta_{k}\gamma_{k}^{-3}\leq0,\label{eq:para_1}
\end{equation}
with $\chi_{k}=\frac{1-\left(\frac{\gamma_{k+1}}{\gamma_{k}}\right)^{2}}{\beta_{k}\gamma_{k}}>0$
as defined in (\ref{eq:chi_w}). By assumptions presented in Theorem~\ref{prop:rate},
$\chi_{k}-A<0$, we can deduce from the inequality (\ref{eq:para_1})
that $\vartheta\geq\overline{\vartheta}_{k}$, with
\[
\vartheta\geq\overline{\vartheta}_{k}=\frac{B}{2\left(A-\chi_{k}\right)}+\sqrt{\left(\frac{B}{2\left(A-\chi_{k}\right)}\right)^{2}+C\frac{\beta_{k}\gamma_{k}^{-3}}{A-\chi_{k}}}.
\]
recall that both $B$ and $C\beta_{k}\gamma_{k}^{-3}$ are positive
by definition. Consider $\epsilon_{1}$ and $\epsilon_{2}$ as defined
in (\ref{eq:small_p}), we have
\begin{align*}
2\overline{\vartheta}_{k} & \leq\frac{B}{A-\epsilon_{1}}+\sqrt{\left(\frac{B}{A-\epsilon_{1}}\right)^{2}+\frac{4C\epsilon_{2}}{A-\epsilon_{1}}},
\end{align*}
We can thus prove that $D_{k+1}\leq\vartheta^{2}\gamma_{k+1}^{2}$
with $\vartheta$ defined in (\ref{eq:para_0}).

Then we turn to prove (\ref{eq:Dk_up2}). Similar to the previous
situation, we have $D_{K_{0}}\leq\varrho^{2}\frac{\beta_{K_{0}}}{\gamma_{K_{0}}}$.
For any $k\geq K_{0}$, if $D_{k}\leq\varrho^{2}\frac{\beta_{k}}{\gamma_{k}}$,
then
\begin{align*}
D_{k+1} & \leq\left(1-A\beta_{k}\gamma_{k}\right)\frac{\beta_{k}}{\gamma_{k}}\varrho^{2}+B\left(\beta_{k}\gamma_{k}\right)^{\frac{3}{2}}\varrho+C\beta_{k}^{2}.
\end{align*}
A sufficient condition to ensure $D_{k+1}\leq\varrho^{2}\frac{\beta_{k+1}}{\gamma_{k+1}}$
is that 
\[
\left(1-A\beta_{k}\gamma_{k}\right)\frac{\beta_{k}}{\gamma_{k}}\varrho^{2}+B\left(\beta_{k}\gamma_{k}\right)^{\frac{3}{2}}\varrho+C\beta_{k}^{2}\leq\frac{\beta_{k+1}}{\gamma_{k+1}}\varrho^{2}.
\]
Thus $\varrho$ should satisfy
\[
\left(\frac{\frac{\beta_{k}}{\gamma_{k}}-\frac{\beta_{k+1}}{\gamma_{k+1}}}{\beta_{k}^{2}}-A\right)\varrho^{2}+B\beta_{k}^{-\frac{1}{2}}\gamma_{k}^{\frac{3}{2}}\varrho+C\leq0
\]
If $\frac{\beta_{k}}{\gamma_{k}}-\frac{\beta_{k+1}}{\gamma_{k+1}}<A\beta_{k}^{2},$
then the condition on $\varrho$ should be $\varrho\geq\overline{\varrho}_{k}$
with
\[
\overline{\varrho}_{k}=\frac{B\beta_{k}^{-\frac{1}{2}}\gamma_{k}^{\frac{3}{2}}+\sqrt{\left(B\beta_{k}^{-\frac{1}{2}}\gamma_{k}^{\frac{3}{2}}\right)^{2}+4C\left(A-\varpi_{k}\right)}}{2\left(A-\varpi_{k}\right)},
\]
where $\varpi_{k}=\beta_{k}^{-2}\left(\frac{\beta_{k}}{\gamma_{k}}-\frac{\beta_{k+1}}{\gamma_{k+1}}\right)>0$
as defined in (\ref{eq:small_p}). Consider $\epsilon_{3}$ and $\epsilon_{4}$
given in (\ref{eq:small_p}), we have
\begin{align*}
\overline{\varrho}_{k} & \leq\frac{B\epsilon_{4}+\sqrt{\left(B\epsilon_{4}\right)^{2}+4C\left(A-\epsilon_{3}\right)}}{2\left(A-\epsilon_{3}\right)},
\end{align*}
then we can prove (\ref{eq:para_0-1}).

\subsection{\label{subsec:Proof-rate_eg}Proof of Theorem~\ref{thm:rate_example}}

From Theorem~\ref{prop:rate}, we can see that the order of the convergence
rate mainly relies on $\nu_{1}$ and $\nu_{2}$, as $\gamma_{k}^{2}\propto\left(k+1\right)^{-2\nu_{2}}$
and $\frac{\beta_{k}}{\gamma_{k}}\propto\left(k+1\right)^{-\left(\nu_{1}-\nu_{2}\right)}$.
However, before the conclusion, we should still verify two points: 
\begin{itemize}
\item \emph{i}) whether the conditions $\epsilon_{1}<A$ and $\epsilon_{3}<A$
are satisfied; 
\item \emph{ii}) whether the constant terms $\vartheta$ and $\varrho$
are bounded. 
\end{itemize}
We start with an elementary and useful result.
\begin{lem}
\label{lem:order}For any $a,b,x\in\left(0,1\right]$, we always have
\[
g\left(x\right)=x^{-a}\left(1-\left(1+x\right)^{-b}\right)<b.
\]
Besides, $\lim_{x\rightarrow0}g\left(x\right)=b$ as $a=1$.
\end{lem}
\begin{IEEEproof}
Since $x^{-a}\leq x^{-1}$ for any $x\in(0,1]$ and $a\in(0,1]$,
we have $g\left(x\right)\leq x^{-1}\left(1-\left(1+x\right)^{-b}\right)=h\left(x\right)$.
We calculate the derivative of $h\left(x\right)$, \emph{i.e.},
\begin{align*}
h'\left(x\right) & =x^{-2}\underbrace{\left(\left(\left(b+1\right)x+1\right)\left(1+x\right)^{-b-1}-1\right)}_{=s(x)}.
\end{align*}
 Thus the monotonicity of $h\left(x\right)$ depends on whether $s(x)$
is positive or negative. We further evaluate 
\begin{align*}
s'(x) & =-b\left(b+1\right)x\left(1+x\right)^{-b-2}\leq0,
\end{align*}
as $b>0$ and $x>0$. Thus $s\left(x\right)$ is a decreasing function
of $x$ over $(0,1]$. It is easy to get $\lim_{x\rightarrow0}s\left(x\right)=0$,
hence, there should be $s\left(x\right)<0$ and $h'\left(x\right)<0$,
$\forall x\in(0,1]$. We have 
\[
h\left(x\right)<\lim_{x\rightarrow0}h\left(x\right)=\frac{1-\left(1+x\right)^{-b}}{x}=b,
\]
which concludes the proof.
\end{IEEEproof}
With the help of Lemma~\ref{lem:order}, we can easily verify $\epsilon_{1}<A$
and $\epsilon_{3}<A$. 
\begin{lem}
\label{lem:condition_verify}Consider $\beta_{k}$ and $\gamma_{k}$
with given forms (\ref{eq:betagamma}), there always exist bounded
$\beta_{0}$ and $\gamma_{0}$ to guarantee $\epsilon_{1}<A$ and
$\epsilon_{3}<A$. 
\end{lem}
\begin{IEEEproof}
Consider (\ref{eq:betagamma}), $\epsilon_{1}$ can be bounded 
\begin{align}
\epsilon_{1} & =\max_{k\geq K_{0}}\frac{1-\left(1+\frac{1}{k+1}\right)^{-2\nu_{2}}}{\beta_{0}\gamma_{0}\left(k+1\right)^{-\nu_{1}-\nu_{2}}}\leq\max_{x\in(0,1]}\frac{1-\left(1+x\right)^{-2\nu_{2}}}{\beta_{0}\gamma_{0}x^{\nu_{1}+\nu_{2}}}\nonumber \\
 & \overset{\left(a\right)}{<}\frac{2\nu_{2}}{\beta_{0}\gamma_{0}},\label{eq:ep1-2}
\end{align}
where $(a)$ is obtained by the application of Lemma~\ref{lem:order}
with $a=\nu_{1}+\nu_{2}$ and $b=2\nu_{2}$. Therefore, we concludes
that $\epsilon_{1}<A$ if $\beta_{0}\gamma_{0}\geq2\nu_{2}/A$. Similarly,
we can show that $\epsilon_{3}<A$ if $\beta_{0}\gamma_{0}\geq\left(\nu_{1}-\nu_{2}\right)/A$,
as
\begin{equation}
\epsilon_{3}=\max_{k\geq K_{0}}\frac{1-\left(1+\frac{1}{k+1}\right)^{-\left(\nu_{1}-\nu_{2}\right)}}{\beta_{0}\gamma_{0}\left(k+1\right)^{-\nu_{1}-\nu_{2}}}<\frac{\nu_{1}-\nu_{2}}{\beta_{0}\gamma_{0}}.\label{eq:ep2}
\end{equation}
\end{IEEEproof}
The remain work is to verify whether the constant term in presence
of the convergence rate can be bounded. Our main result stated in
Theorem~\ref{thm:rate_example} can be easily obtained.

We start with the analysis of $\epsilon_{2}$ and $\epsilon_{4}$,
\emph{i.e.},
\begin{align*}
\epsilon_{2} & =\beta_{0}\gamma_{0}^{-3}\max_{k\geq K_{0}}\left(1+k\right)^{-\left(\nu_{1}-3\nu_{2}\right)}\\
 & =\begin{cases}
\beta_{0}\gamma_{0}^{-3}\left(1+K_{0}\right)^{-\left(\nu_{1}-3\nu_{2}\right)}, & \textrm{if }\nu_{1}\geq3\nu_{2},\\
\infty, & \textrm{if }\nu_{1}<3\nu_{2},
\end{cases}
\end{align*}
and 
\begin{align*}
\epsilon_{4} & =\beta_{0}^{-\frac{1}{2}}\gamma_{0}^{\frac{3}{2}}\max_{k\geq K_{0}}\left(1+k\right)^{\frac{\nu_{1}-3\nu_{2}}{2}}\\
 & =\begin{cases}
\beta_{0}^{-\frac{1}{2}}\gamma_{0}^{\frac{3}{2}}\left(1+K_{0}\right)^{\frac{\nu_{1}-3\nu_{2}}{2}}, & \textrm{if }\nu_{1}\leq3\nu_{2},\\
\infty, & \textrm{if }\nu_{1}>3\nu_{2}.
\end{cases}
\end{align*}
We can see that $\epsilon_{2}$ and $\epsilon_{4}$ cannot be bounded
under the same condition, unless $\nu_{1}=3\nu_{2}$. 

When $\nu_{1}>3\nu_{2}$, $\epsilon_{2}$ is bounded, we can say that
$\vartheta$ is also bounded by its definition, as long as $\beta_{0}\gamma_{0}\geq2\nu_{2}/A$
(recall Lemma~\ref{lem:condition_verify}). Meanwhile, $\varrho\rightarrow\infty$
as $\epsilon_{4}\rightarrow\infty$, which makes the bound (\ref{eq:Dk_up2})
loose. Therefore, there exists some bounded constant $\Omega_{1}$,
such that $D_{k}\leq\Omega_{1}\left(k+1\right)^{-2\nu_{2}}$.

When $\nu_{1}<3\nu_{2}$, $\epsilon_{4}$ is bounded whereas $\epsilon_{2}\rightarrow\infty$.
Then there exists $\Omega_{2}<\infty$, such that $D_{k}\leq\Omega_{2}\left(k+1\right)^{-\left(\nu_{1}-\nu_{2}\right)}$
if $\beta_{0}\gamma_{0}\geq\left(\nu_{1}-\nu_{2}\right)/A$.

When $\nu_{1}=3\nu_{2}$, both $\epsilon_{2}$ and $\epsilon_{4}$
are bounded, the similar result can be obtained, which concludes the
proof.

\bibliographystyle{IEEEtran}
\bibliography{BiblioWenjie}

\end{document}